\documentclass{jfm}
\usepackage{graphicx}
\usepackage{epstopdf, epsfig}
\usepackage{amsmath}
\usepackage{mathtools}

\shorttitle{Convergence of the normal form transformation}
\shortauthor{S. G. Walsh and M. D. Bustamante}

\title{On the convergence of the normal form transformation in discrete Rossby and drift wave turbulence }

\author{Shane G. Walsh\aff{1}
  \corresp{\email{shanewalsh11235@gmail.com}}
 \and Miguel D. Bustamante\aff{1} }

\affiliation{\aff{1}School of Mathematics and Statistics, University College Dublin, Belfield,
  Dublin 4, Ireland
}

\begin{document}

\maketitle

\begin{abstract}
We study numerically the region of convergence of the normal form transformation for the case
of the Charney-Hasagawa-Mima (CHM) equation to investigate whether certain finite amplitude 
effects can be described in normal coordinates. We do this by taking a Galerkin truncation of
four Fourier modes making part of two triads: one resonant and one non-resonant, joined together
by two common modes. We calculate the normal form transformation directly from the 
equations of motion of our reduced model, successively applying the algorithm to calculate
the transformation up to $7^{\mathrm{th}}$ order to eliminate all non-resonant terms, and keeping up to
8-wave resonances. We find that the amplitudes at which the normal form transformation 
diverge very closely match with the amplitudes at which  
a finite-amplitude phenomenon called \emph{precession resonance}~\citep{Busta} occurs, characterised by strong energy transfers. This
implies that the precession resonance mechanism cannot be explained using the usual methods
of normal forms in wave turbulence theory, so a more general theory for intermediate 
nonlinearity is required.

\end{abstract}

\section{Introduction}
\label{sec:intro}
Many real-life systems of engineering and physical interest are governed by nonlinear wave
equations: nonlinear circuits, nonlinear optics, ocean surface waves, planetary waves in 
the atmosphere,  etc.~\citep{rewienski2003,Kibler2010,Kraichnan,lynch2009} One of the 
hallmarks of nonlinear dynamics is the strong transfer of energy across scales, 
so modelling and understanding energy transfer in nonlinear systems is key to 
predicting and harnessing extreme events in these physical systems.  One such system, the 
Charney-Hasagawa-Mima (CHM) equation (which will be the focus of this paper) 
is a nonlinear dispersive partial differential equation which can be used to describe Rossby
waves in the atmosphere and drift waves in plasmas. 

One approach to understanding how nonlinear systems behave is model reduction, 
where one simplifies the system to a point that is easier to study mathematically, while leaving
enough complexity in the system so that the phenomena under scrutiny remain in the reduced
model. A common method is spectral truncation, where one considers only a small set of 
interacting modes in Fourier space. Often these reduced models are studied from a 
dynamical systems point of view where interesting feedback mechanisms have been proposed~\citep{waleffe1997}: as more complex models are considered, these mechanisms have survived in the form of periodic orbits. Other dynamical systems approaches in fluid mechanics include
studying shell models~\citep{biferale2003,mailybaev2013}, 
and the search and classification of unstable periodic orbits 
(UPOs) in higher dimensional systems~\citep{kawahara2001, lucas2015,lucas2017}. 
A useful method for analysing models that are based on dynamical systems of any dimension is the method of normal forms. The normal form transformation of a dynamical system is often described as a nonlinear 
transformation of coordinates in which the dynamical system takes its ``simplest'' 
form~\citep{wiggins}. There is a certain amount of freedom one can take with the form of the
transformation, however its usual manifestation involves eliminating as many non-resonant
terms as possible from the evolution equations. This method dates back to Poincar\'e, and was
developed for the $N$-body problem in celestial mechanics. Analytic study of convergence
of these normal form transformations is quite difficult and often concerns with whether the
series will converge at all \citep{bruno2011local}. Very little work has been done on studying
the region of convergence of these transformations. For example, in the FPUT system the 
normal form transformation is important to proving integrability at small amplitudes and to establishing the applicability of Kolmogorov-Arnold-Moser theorem, however as explained by \cite{Rink2006} no study of convergence has been 
performed to date.

Despite the lack of rigorous convergence results, one of the most fruitful approaches to studying nonlinear wave equations has been the 
theory of wave turbulence \citep{nazarenko2011wave}, which uses normal form transformations to eliminate non-resonant
interactions from the system. In the usual formulation of wave turbulence, the limits of both
small amplitudes and large domain are considered. The small amplitudes allow for the non-resonant
terms to be eliminated as the non-zero linear frequencies of oscillation in the system are
much faster than the 
nonlinear frequencies, and the large box limit allows for a continuum of wave numbers in the 
system. For some systems these limits are a good approximation. An example of this is gravity
water waves, where the steepness of the waves is typically small and a large domain (compared to the relevant wavelengths) makes
physical sense. Normal form transformations have been very effective in this case: these transformations were originally calculated by Zakharov and 
Dyachenko~\citep{dyach} in a formal (i.e. non-rigourous) way, but  recently these transformations were shown to be rigorously correct, leading to integrability up to fourth 
order along with long-time existence results~\citep{berti2018}.

However, all physical systems have finite amplitude and all numerical simulations
must be performed on a finite domain, and moreover there are many nonlinear wave systems where
the large-box or weak-amplitude approximations are not good assumptions, so studying the behaviour of systems when 
these two assumptions are relaxed is important. Once finite amplitudes are introduced it is 
necessary to understand how and where the normal form transformation converges.

In this paper we wish to study the region of convergence of the normal form transformation 
for a truncated system that does not necessarily follow from a Hamiltonian principle. 
 In Section~\ref{sec:precres}
we describe the CHM equation and how at intermediate Fourier amplitudes the so-called precession
resonance~\citep{Busta} can be observed. We then take a low-dimensional Galerkin truncation 
that demonstrates precession resonance and investigate how the resonance manifests itself in
state space. In Section~\ref{sec:transf} we define the normal form transformation and detail how it is 
developed out of the evolution equations of our truncated system. The usual wave turbulence approach
involves calculating the normal form transformation from a Hamiltonian; however in the case
of the CHM equation it is more natural to compute it directly from the evolution equations.
From there we calculate the 
transformation eliminating up to non-resonant 4-wave interactions and study the system
analytically. In Section~\ref{sec:pwc} we continue calculating the transformation to higher orders, eliminating
up to non-resonant 8-wave interactions and then calculate the rate of convergence of the
transformation by performing linear regression on the size of the terms with increasing order.
From here we numerically find the amplitude at which the transformation begins to diverge.
We compare this with the amplitude where precession resonance occurs and we find that there is 
a strong connection between the divergence of the transformation and precession resonance. Finally, we provide concluding remarks in Section~\ref{sec:concl}.

\section{Precession Resonance in the CHM equation}
\label{sec:precres}
\subsection*{The CHM equation and the weakly nonlinear limit}
Consider the Charney Hasagawa Mima (CHM) equation, a PDE model for Rossby waves in the atmosphere as well as drift waves in plasmas:
\begin{equation}
  (\nabla^2-F)\frac{\partial \psi}{\partial t} + \beta \frac{\partial \psi}{\partial x} + \frac{\partial \psi}{\partial x}\frac{\partial\nabla^2 \psi}{\partial y} - \frac{\partial \psi}{\partial y}\frac{\partial\nabla^2 \psi}{\partial x} = 0
\end{equation}

where $\psi$ is the streamfunction of a geophysical flow and $F\geq 0$ is the Rossby
deformation radius. Assuming periodic boundary conditions $\mathbf{x}\in[0,2\pi)^2$, we can
decompose the equation into its Fourier components $A_\mathbf{k}$ defined by 
$\psi(\mathbf{x},t)=\sum_{\mathbf{k}\in\mathbb{Z}^2}A_{\mathbf{k}}(t)e^{i \mathbf{k\cdot x}}+\text{c.c.}$, which leads to

\begin{equation}
  \dot{A}_\mathbf{k} + i \omega_\mathbf{k} A_\mathbf{k} = \frac{1}{2}\sum_{\mathbf{k}_1\mathbf{k}_2\in\mathbb{Z}}Z^\mathbf{k}_{\mathbf{k}_1\mathbf{k}_2}\delta_{\mathbf{k}_1+\mathbf{k}_2-\mathbf{k}}A_{\mathbf{k}_1}A_{\mathbf{k}_2}\,, \qquad  \mathbf{k}\in \mathbb{Z}^2\,,
  \label{eq:chmfourier}
\end{equation}

where 
\begin{equation*}
  \begin{split}
    Z^\mathbf{k}_{\mathbf{k}_1\mathbf{k}_2} &= \frac{(k_{1 x}k_{2 y}- k_{1 y} k_{2 x})(|\mathbf{k}_1|^2-|\mathbf{k}_2|^2)}{|\mathbf{k}|^2 + F}, \qquad \omega_{\mathbf{k}} = \frac{-\beta k_x}{|\mathbf{k}|^2 + F} \,.
  \end{split}
\end{equation*}
\\
Due to the Kronecker delta on the right hand side of equation~(\ref{eq:chmfourier}), we can see
that the only nonlinear contributions to mode $A_\mathbf{k}$ come from modes $A_{\mathbf{k}_1}$
and $A_{\mathbf{k}_2}$ whenever $\mathbf{k} = \mathbf{k}_1+\mathbf{k}_2$. These kinds of
interactions are known as triad interactions \citep{Kraichnan}.\\

In the limit of small amplitudes, the evolution of $A_\mathbf{k}$ in 
equation~(\ref{eq:chmfourier}) is dominated by the linear dispersion relation 
$\omega_\mathbf{k}$. These waves are known as Rossby waves.
To see how the nonlinear term contributes to the dynamics we perform a change of variable
 $a_\mathbf{k}(t)=e^{i\omega_{\mathbf{k} }t}A_{\mathbf{k}}(t)$ to obtain our
equations in the so-called interaction representation:

\begin{equation}
  \dot{a}_\textbf{k} = \frac{1}{2}\sum\limits_{\textbf{k}_1, \textbf{k}_2 \in \mathbb{Z}^2} Z^\textbf{k}_{\textbf{k}_1 \textbf{k}_2}\, \delta_{\textbf{k}_1 +\textbf{k}_2-\textbf{k}}\, a_{\textbf{k}_1}\, a_{\textbf{k}_2}e^{-i \omega^{\mathbf{k}}_{\mathbf{k}_1 \mathbf{k}_2} t} \,.
  \label{eq:interact}
\end{equation}

where $\omega^{\mathbf{k}}_{\mathbf{k}_1 \mathbf{k}_2}=\omega_{\mathbf{k}}- \omega_{\mathbf{k}_1}-\omega_{\mathbf{k}_2}$.

Inspecting equation~(\ref{eq:interact}) we can see that when the amplitudes $a_\mathbf{k}$ are small (so-called weakly nonlinear limit)
they become slow as well.
 This implies that the fast oscillations of 
$e^{-i \omega^{\mathbf{k}}_{\mathbf{k}_1 \mathbf{k}_2} t}$ average out to zero for
$\omega^{\mathbf{k}}_{\mathbf{k}_1 \mathbf{k}_2} \neq 0$, meaning that the only meaningful
triad interactions occur when $\omega_{\mathbf{k}}=\omega_{\mathbf{k}_1}+\omega_{ \mathbf{k}_2}$, the so-called resonant condition. 
Thus in the weakly nonlinear limit the non-resonant triad interactions (defined by the inequality
$\omega^{\mathbf{k}}_{\mathbf{k}_1 \mathbf{k}_2}\neq0$) do not contribute 
to the long-time dynamics of the system. In the classical theory of wave turbulence 
a nonlinear near-identity change of coordinates is preformed to eliminate these non-resonant interactions from
the equations, in what is often referred to as a normal form transformation \citep{nazarenko2011wave}.

\subsection*{Precession resonance}
To understand precession resonance we must first consider the equations in phase-amplitude
form, where $$A_\mathbf{k} = n_{\mathbf{k}}e^{i\phi_\mathbf{k}}$$
where $n_\mathbf{k}\in[0,\infty)$ and $\phi_\mathbf{k}\in[0,2\pi).$ To allow for phase precessions \citep{Busta} (also known as phase-slips \citep{Knobloch}) we will consider $\phi_\mathbf{k}\in (-\infty, \infty)$.
Upon transforming our evolution equations to phase-amplitude form we will see that due to the triad interactions the Fourier phases do not appear isolated.  Rather, they 
appear in triad combinations
$\varphi^\mathbf{k}_{\mathbf{k}_1 \mathbf{k}_2}=\phi_{\mathbf{k}_1}+\phi_{\mathbf{k}_2}-\phi_{\mathbf{k}}$ known as Fourier triad phases, where $\mathbf{k}_1 + \mathbf{k}_2 =  \mathbf{k}$. Our equations take the form
\begin{align}
  \dot{n}_{\mathbf{k}}&= \sum_{\mathbf{k}_1 \mathbf{k}_2}Z^\mathbf{k}_{\mathbf{k}_1\mathbf{k}_2}\delta_{\mathbf{k}_1+\mathbf{k}_2-\mathbf{k}}n_{\mathbf{k}_1}n_{\mathbf{k}_2}\cos\varphi^\mathbf{k}_{\mathbf{k}_1 \mathbf{k}_2} \,,\label{eq:amp}\\ 
    \dot{\varphi}^{\mathbf{k}_3}_{\mathbf{k}_1 \mathbf{k}_2}&=\sin\varphi^{\mathbf{k}_3}_{\mathbf{k}_1 \mathbf{k}_2}n_{\mathbf{k}_3}n_{\mathbf{k}_1}n_{\mathbf{k}_2}\left(\frac{Z^{\mathbf{k}_1}_{\mathbf{k}_2\mathbf{k}_3}}{n_{\mathbf{k}_1}}+\frac{Z^{\mathbf{k}_2}_{\mathbf{k}_3\mathbf{k}_1}}{n_{\mathbf{k}_2}}-\frac{Z^{\mathbf{k}_3}_{\mathbf{k}_1\mathbf{k}_2}}{n_{\mathbf{k}_3}} \right) - \omega^{\mathbf{k}_3}_{\mathbf{k}_1 \mathbf{k}_2} + \text{NNTT}^{\mathbf{k}_3}_{\mathbf{k}_1 \mathbf{k}_2}. \label{eq:phase}
  \end{align}

The term $\text{NNTT}^{\mathbf{k}_3}_{\mathbf{k}_1 \mathbf{k}_2}$ are the nearest neighbouring
triad terms connected to the triad $\mathbf{k}_1+\mathbf{k}_2=\mathbf{k}_3$. This term can be seen in full in \citep{Busta}.

We define precession frequency as $\Omega^\mathbf{k}_{\mathbf{k}_1 \mathbf{k}_2} \equiv \lim_{t \to \infty}(1/t)\int_0^t\dot{\varphi}^\mathbf{k}_{\mathbf{k}_1 \mathbf{k}_2}(t')\mathrm{d}
t'$.
Geometrically this corresponds to the frequency at which $\varphi^\mathbf{k}_{\mathbf{k}_1 \mathbf{k}_2}$ winds around the origin.  
In the weakly nonlinear case this precession frequency is dominated by the linear term, but
if we extend our system beyond the weakly nonlinear limit and start taking into account finite
amplitudes, the nonlinear terms begin to contribute to the dynamics and in particular to the precession frequency.

Looking at the right hand side of equation~(\ref{eq:amp}), we can see that if the characteristic
nonlinear freqency of $n_{\mathbf{k}_1}n_{\mathbf{k}_2}$ is commensurate with the 
precession frequency $\Omega^\mathbf{k}_{\mathbf{k}_1 \mathbf{k}_2}$
we then get a zero mode in the evolution equation of $n_\mathbf{k}$ which leads to sustained 
growth. The strongest manifestation of this is the zero harmonic resonance, where 
$\Omega^\mathbf{k}_{\mathbf{k}_1 \mathbf{k}_2}=0.$ It is clear from equation~(\ref{eq:amp}) that
if this condition is satisfied then we obtain strong growth in mode $A_{\mathbf{k}}$. One of the 
remarkable things about precession resonance is that we can always trigger this resonance simply
by rescaling the initial conditions by a constant $\alpha$. As we increase $\alpha$, the 
nonlinear contributions to equation~(\ref{eq:phase}) can become  of the order of the 
linear part $\omega^{\mathbf{k}_3}_{\mathbf{k}_1 \mathbf{k}_2}$ and cancel, resulting in
$\Omega^\mathbf{k}_{\mathbf{k}_1 \mathbf{k}_2}=0.$

\subsection*{Model reduction}

\begin{figure}[h]
     \centering
       \includegraphics[width=0.44\textwidth]{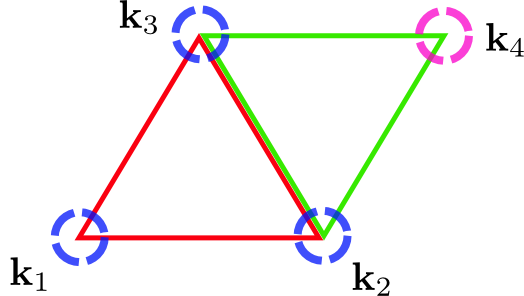}
       \includegraphics[width=0.5\textwidth]{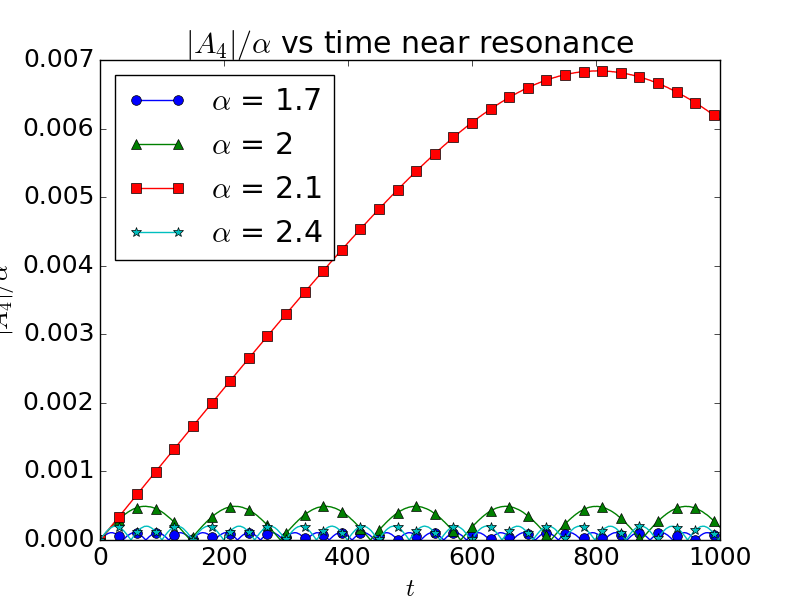}
      
     \caption{Left panel: Schematic representation of our reduced system. Right panel: Plots of the normalised function of time $|A_4(t)|/\alpha$, for initial conditions from equation (\ref{eq:init_conds}) in terms of the scaling parameter $\alpha$ taking different values as in the legend. The resonant scaling value at $\alpha = \alpha_r\approx 2.11$ shows a strong energy transfer to $A_4$.
     }
   \label{fig:4mode}
   \end{figure}

The simplest manifestation of precession resonance for the CHM model occurs in a Galerkin truncation to four Fourier
modes. Our system is composed of a resonant triad joined to a fourth mode 
by two modes via a non-resonant triad (see figure \ref{fig:4mode}). Putting a small parameter $\epsilon$
in front of the second triad interaction coefficients allows us to match our results with the reduced model in 
\citep{Busta}. It also allows us to focus our attention on the effect of precession resonance on the energy transfers to one 
particular mode, which we choose to be $A_4$.\\
Our equations are:

\begin{equation}
  \begin{split}
    \dot{A}_{1}&=-i \omega_{1}A_{1} + z_1 A_{2}^*A_{3}\\
    \dot{A}_{2}&=-i \omega_{2}A_{2} + z_2 A_{1}^*A_{3} +\epsilon s_2 A_{3}^*A_{4}\\
    \dot{A}_{3}&=-i \omega_{3}A_{3} + z_3 A_{1}A_{2} +\epsilon s_3 A_{2}^*A_{4}\\
    \dot{A}_{4}&=-i \omega_{4}A_{4} +\epsilon s_4 A_{2}A_{3}\\
    \omega_3 &= \omega_1 +\omega_2\text{,} \quad \omega_4 \neq \omega_2 +\omega_3\,,
  \end{split}
  \label{eq:4mode}
\end{equation}

our parameters being $F=1$, $\beta=10$, $\epsilon = 0.01$, 
$\mathbf{k}_1 = (1,-4)$, $\mathbf{k}_2 = (1,2)$, $\mathbf{k}_3 = (2,-2)$, 
and $\mathbf{k}_4 = (3,0)$, $\omega^{\mathbf{k}_3}_{\mathbf{k}_1 \mathbf{k}_2} = 0 $ and $\omega^{\mathbf{k}_4}_{\mathbf{k}_2 \mathbf{k}_3} = -8/9$.

In order to search for precession resonance we will explore a set of initial conditions for our variables $A_j, \,\, j=1, \ldots, 4$, of the form
\begin{equation}
\label{eq:init_conds}
A_1(0) = (0.0245 + 0.001i)\alpha, \quad A_2(0) = (0.01 + 0.01i)\alpha, \quad  A_3(0) =0.02236\alpha, \quad A_4(0) = 0,
\end{equation}
where, except for $A_4(0)$, the numerical coefficients were randomly chosen, and $\alpha$ is our real scaling
parameter. The size of the numerical coefficients was chosen so that the linear terms in the equations dominate over 
the nonlinear terms when $|\alpha| \ll 1$. 

In figure~\ref{fig:4mode} on the right we observe that as we get close to the exact scaling value ($\alpha_r \approx 2.11$) for the initial condition (\ref{eq:init_conds}) which leads to precession
resonance, our energy transfer to mode $A_4$ greatly increases. As we scale past that value, 
the energy transfer efficiency becomes weak again.\\
The choice of initial conditions with $A_4(0)=0$ is
for visualisation purposes as well as for maximising the effect of precession resonance to this 
mode. Precession resonance can occur for any arbitrary initial condition (see the study shown on the left panel of figure \ref{fig:gamma}). However, a
particularly interesting facet of precession resonance is the prospect of energy leaking to
modes which start off with no energy and are not part of any resonant triads.

This phenomenon was described and studied in \citep{Busta}. In this document we wish to study
the resonance in normal-form coordinates used in wave turbulence theory to see if the resonance
can manifest itself in the dynamics for the normal-form coordinates. We also wish to better
understand the geometric structure of the resonance in state space.

   \begin{figure}[h]
     \centering
       \includegraphics[width=0.8\textwidth]{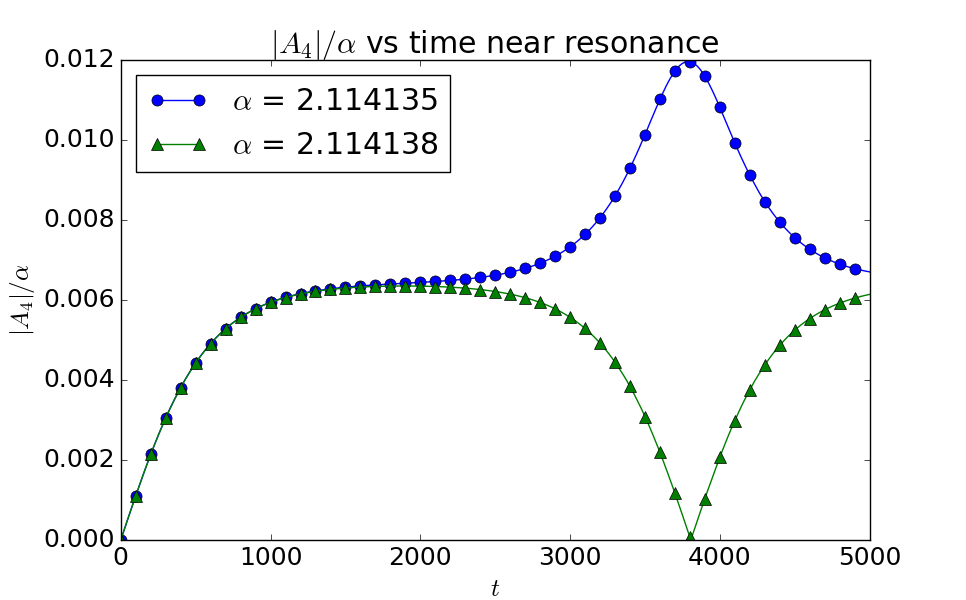}
       \caption{For $\alpha\approx\alpha_r \pm 10^{-6}$ saddle node like behaviour can be seen around the trajectory associated with precession resonance.}
       \label{fig:xpoint}
   \end{figure}

\subsection*{Invariant Manifolds}
\label{sec:invman}

For near resonant values of our scaling parameter $\alpha$, figure~\ref{fig:xpoint} shows saddle-node-like behaviour. This suggests a dynamical systems point of view whereby resonant trajectories in
the reduced model could correspond to invariant manifolds such as critical points or periodic orbits in state space.

\begin{figure}
     \centering
       \includegraphics[width=0.5\textwidth]{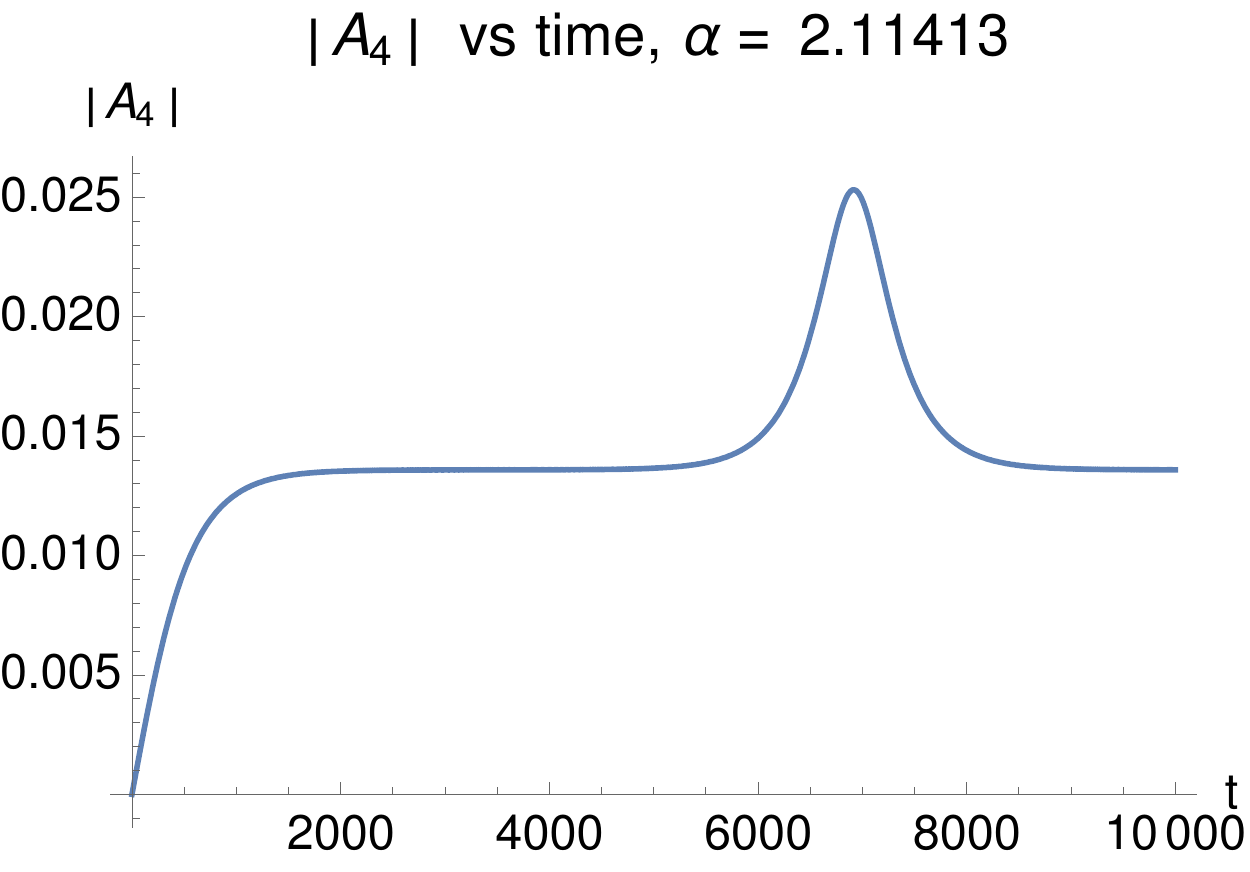}
\hfill
       \includegraphics[width=0.45\textwidth]{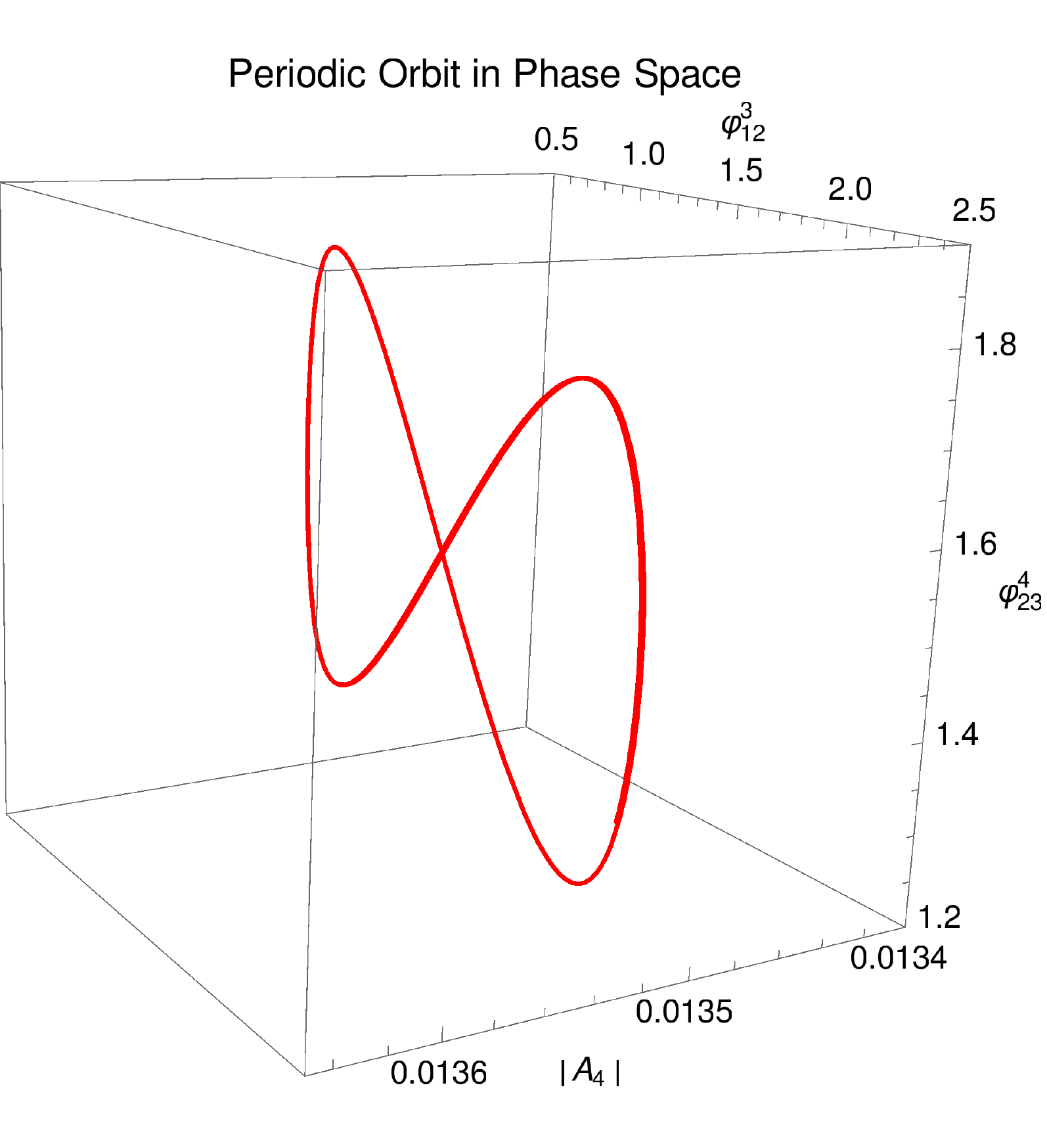}
   \caption{Left panel: For $\alpha\approx\alpha_r$, $A_4$ approaches a particular value. After a time it is ejected along an unstable manifold. Right panel: Zooming into a projection of the trajectory onto $|A_4|$, $\varphi^{3}_{12}$ and $\varphi^{4}_{23}$, we can see that the orbit approaches a periodic orbit. }
   \label{fig:longsimorb2}
   \end{figure}

Figure~\ref{fig:longsimorb2} shows that, letting this system evolve for a long time, the resonance corresponds to a trajectory that gets close to a periodic orbit in state space and remains close for a while to then separate from it. 

To gain more insight into the geometric structure of this resonance, we consider how these 
manifolds manifest themselves in the state space. First, let us count the number of degrees of freedom. Equations~(\ref{eq:4mode}) represent 4 complex equations for 4 complex
variables. However, in terms of amplitude-phase representation, the phases appear only in two triad combinations so we can reduce the state space to 
6 dynamical variables. Second, we have two conserved quantities: energy and enstrophy, defined as:
\begin{equation}
  \begin{split}
    E =& \sum_{i=1}^4(|\mathbf{k}_i|^2+F)|A_i|^2 \\
    \mathcal{E} =&\sum_{i=1}^4  |\mathbf{k}_i|^2(|\mathbf{k}_i|^2+F)|A_i|^2
  \end{split}
  \label{eq:energy}
\end{equation}

which further reduces the dimension of the system by 2 units, leading to an effective 4-dimensional system.

In the previous Section we searched for resonances by re-scaling the initial conditions by a 
constant. While this shows how simple it is to find these resonances, it doesn't shed any light
on the structure of the periodic orbits in state space. This prompts us to search for resonances
contained in the invariant manifolds corresponding to fixed energy and enstrophy.

To perform this search of
resonances constrained to fixed energy and enstrophy, we first proceed using the same re-scaling technique as before with $\alpha$ until we trigger a resonance.
For the same initial conditions we used in Section~\ref{sec:precres} we found that resonance
occured at $\alpha_r \approx 2.114 $ giving resonant initial conditions to be $\mathbf{A}\approx(0.0518+0.0021 i,\, 0.0211 + 0.0211 i,\, 0.0472,\, 0 )^T$.

\begin{figure}[h]
     \centering
  \includegraphics[width=0.5\textwidth]{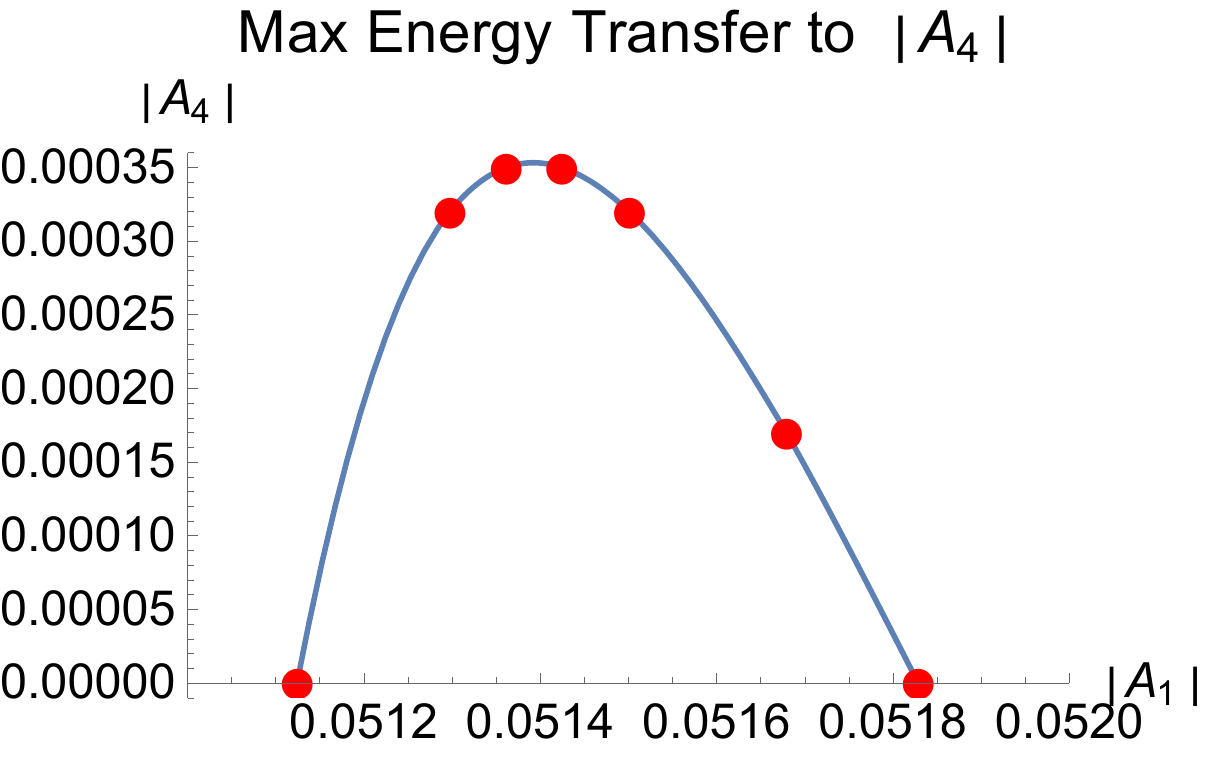}
     \includegraphics[width=0.45\textwidth]{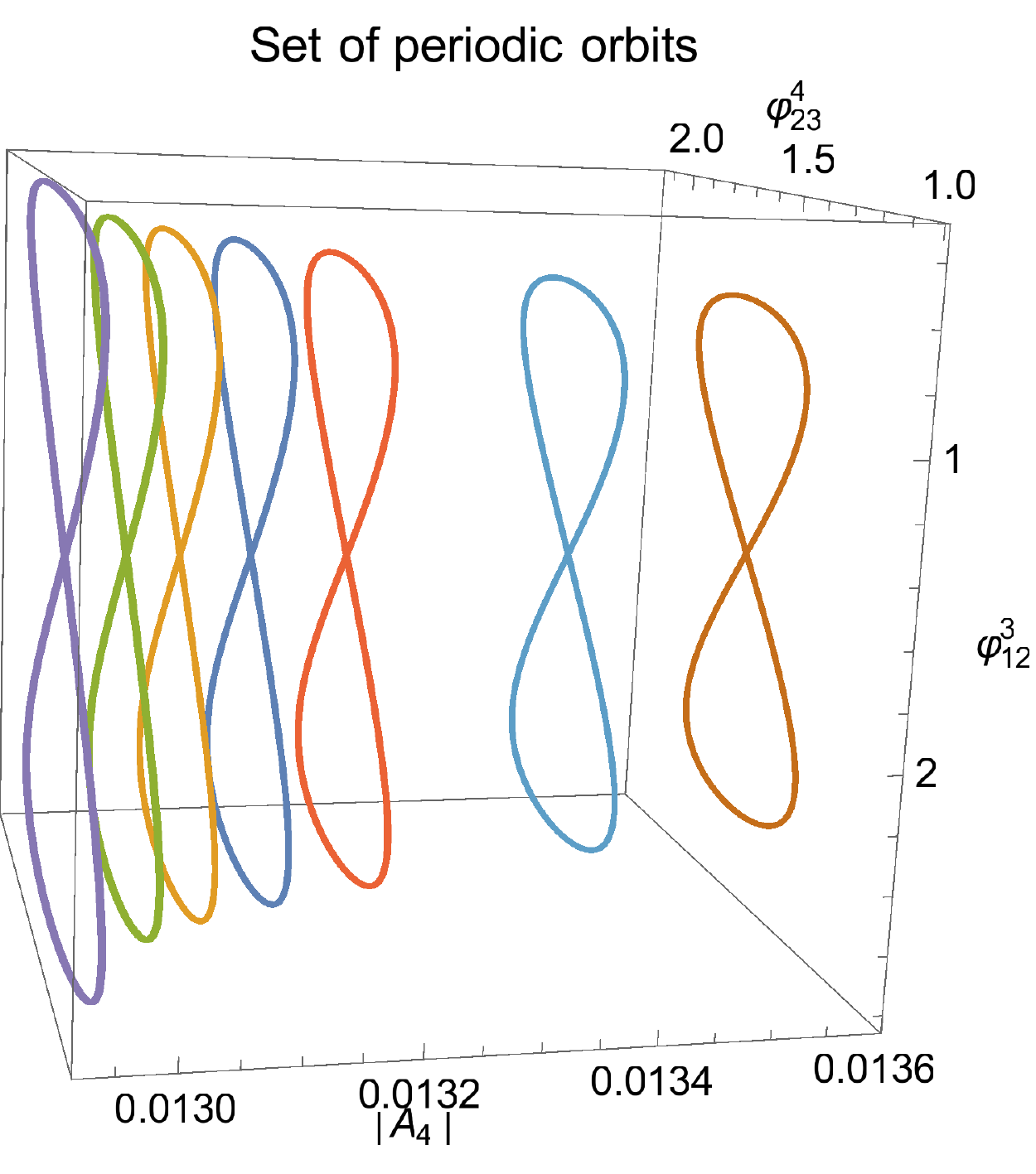}
\caption{Left panel: A curve of initial values for $|A_1|$ and $|A_4|$ on the same invariant
  energy and enstrophy manifolds that lead to precession resonance. The points correspond to 
  the periodic orbits plotted in the right panel, in the sense that the initial condition implied by each point belongs to the stable manifold of the corresponding periodic orbit. Right panel: A set of periodic orbits associated with
  precession resonance on the same invariant manifolds. The orbits from right to left correspond
  to the points from right to left on the left panel.}
   \label{fig:gamma}
   \end{figure}

From these initial values we can now calculate the values for energy $E=0.0738$ and enstrophy $\mathcal{E}=1.0097$ from equation~(\ref{eq:energy}). We will fix these values and perform  a systematic search for resonances within the intersection of the manifolds $E=0.0738$ and $\mathcal{E}=1.0097$.
To start, choose as initial condition $|A_{4\,\text{new}}|=|A_{4\,\text{old}}|+\delta$ where $\delta$ is a small positive number. We now vary the initial $|A_1|$, solving for $|A_2|$ and $|A_3|$ to make sure $E$ and $\mathcal{E}$ are unchanged, until a new resonance point is found.
Throughout these searches we keep the initial complex phases unchanged. It was found that 
changing the complex phases still lead to a periodic orbit that could already be found using
our current method, it would just shift the initial condition to another location along the orbit.
As for the phase of the initial $A_4$, although our first initial value was $A_4=0$ and so $\arg(A_4)$ was undefined, we have picked an arbitrary
value for $\arg(A_4)$  when $A_4\neq0$ (in our case we chose $\arg(A_4)=0$).
 The systematic search produces a continuous curve representing the points of resonance as can be seen in figure \ref{fig:gamma}, left panel.

Figure \ref{fig:gamma} (right panel) shows the periodic orbits corresponding to some resonance points chosen along the curve in Figure \ref{fig:gamma} (left panel). So the resonant curve corresponds to a one-parameter family of periodic orbits. It is evident that two directions in state space have zero Lyapunov exponent: the direction along the periodic-orbit time evolution and the direction that connects the different periodic orbits. Since our system has 4 degrees of freedom
and we already need to have a stable and an unstable manifold to reach the periodic orbits, we conclude that we can determine the structure and dimensions
of all relevant manifolds in state space. We have a one-dimensional stable manifold, a one-dimensional unstable manifold, and two neutral directions induced by the one-dimensional time evolution along the periodic orbits and the one-dimensional direction along which the periodic orbits are ordered. Moreover, from the fact that the original system is volume-preserving, the Lyapunov exponents of the stable and unstable manifolds are $-\Lambda$ and $\Lambda$, respectively, with $\Lambda>0$.

This gives us a clearer picture of precession resonance from a dynamical systems point of view
for this 4-mode model. Precession resonance occurs when our initial condition is close to the stable
manifold of a periodic orbit far from the origin.

\section{The Normal Form Transformation}
\label{sec:transf}
We have seen in our 4-mode system that precession resonance can be understood as a resonance
between the linear and nonlinear oscillations of the system.In classical wave turbulence
theory a scale separation is assumed between linear and nonlinear timescales so that non-resonant terms are eliminated from the equations through normal form 
transformations. This raises the questions: Can precession resonance manifest itself
in normal-form coordinates? If yes, how does it do it?
For Hamiltonian wave systems, a canonical transformation for eliminating non-resonant $n$-wave
interactions has been well described in \citep{krasitskii} and \citep{dyach}. However, as 
the Hamiltonian structure of the CHM equation requires both a non-canonical Poisson bracket and a non-local transformation of coordinates~\citep{awein}, it is both easier and more illustrative to calculate the normal form out of the evolution equations directly.

We will use the method described in~\citep{wiggins}.
We start with a system of the form
$$ \dot{\mathbf{A}} = J\mathbf{A} + F(\mathbf{A})$$
where $J$ is a matrix with constant valued elements and
$$F(\mathbf{A})=F^{(2)}(\mathbf{A})+F^{(3)}(\mathbf{A})+\ldots$$
where $F^{(n)}(\mathbf{A})$ is a vector valued homogeneous polynomial of degree $n$. 
To eliminate the second order nonlinearities, we perform a near identity 
transformation of the form $$\mathbf{A} = \mathbf{B} + h^{(2)}(\mathbf{B})$$
where $h^{(2)}$ is an unknown function quadratic in components of $\mathbf{B}$. Subbing this into the
original equation we get
$$\dot{\mathbf{B}} = J \mathbf{B} + J h^{(2)}(\mathbf{B}) - \nabla h^{(2)}(\mathbf{B})J \mathbf{B} + F^{(2)}(\mathbf{B}) + \tilde{F}^{(3)}(\mathbf{B})+\ldots$$
    
In an ideal situation we would choose our $h^{(2)}$ such that
$$ F^{(2)}(\mathbf{B})=\nabla h^{(2)}(\mathbf{B})J \mathbf{B}-J h^{(2)}(\mathbf{B})$$
however with the inclusion of resonant terms we cannot make such a choice. We can however
choose $h^{(2)}(\mathbf{B})$ such that 
$$ J h^{(2)}(\mathbf{B}) - \nabla h^{(2)}(\mathbf{B})J\mathbf{B} + F^{(2)}(\mathbf{B}) = R^{(2)}(\mathbf{B})$$
leaving just the resonant terms and eliminating all other terms of that order, leaving higher
order corrections.

This leaves us with a new system of equations for $\mathbf{B}$ of the form:
$$\dot{\mathbf{B}} = J \mathbf{B} + R^{(2)}(\mathbf{B}) + \tilde{F}^{(3)}(\mathbf{B})+\ldots\\$$
where $R^{(n)}(\mathbf{B})$ are the resonant terms of order $n$ (i.e. $(n+1)$-wave resonant 
interactions). We can continue this transformation and eliminate higher order interactions via

$$\mathbf{B}=\mathbf{C} + h^{(3)}(\mathbf{C}) + \ldots$$
leading to 
$$\dot{\mathbf{C}} = J \mathbf{C} + R^{(2)}(\mathbf{C}) + R^{(3)}(\mathbf{C})+\ldots.$$
Analogously to the small-amplitude, continuum formulation of wave turbulence, we now 
transform equation (\ref{eq:4mode}) using our method in order to eliminate the non-resonant nonlinearities, leaving corrections at the next order.  

As a preliminary step we can just eliminate the non-resonant triads. The required transformation is
\begin{equation}
  \begin{split}
  A_1 &= B_1 \\
  A_2 &= B_2 - \frac{i s_2 B_3^* B_4}{\omega_2 +\omega_3 -\omega_4}\\
  A_3 &= B_3 - \frac{i s_3 B_2^* B_4}{\omega_2 +\omega_3 -\omega_4}\\
  A_4 &= B_4 + \frac{i s_4 B_2 B_3}{\omega_2 +\omega_3 -\omega_4}\\
  \end{split}
\end{equation}

and the corresponding equations of motion for the new variables are:
\begin{equation}
  \begin{split}
  \dot{B}_{1}&=-i \omega_{1}B_{1} + z_1 B_{2}^*B_{3} + \mathcal{O}(|\mathbf{B}|^3) \qquad \\
  \dot{B}_{2}&=-i \omega_{2}B_{2} + z_2 B_{1}^*B_{3} + \mathcal{O}(|\mathbf{B}|^3)\\
  \dot{B}_{3}&=-i \omega_{3}B_{3} + z_3 B_{1}B_{2} + \mathcal{O}(|\mathbf{B}|^3)\\
  \dot{B}_{4}&=-i \omega_{4}B_{4} + \mathcal{O}(|\mathbf{B}|^3)\\
  \end{split}
\end{equation}

If we substitute $B_j$ in terms of the interaction representation, $ b_j = B_j e^{i \omega_{j} t}, \,\, j=1, \ldots, 4$, we see that the equations reduce to the isolated resonant triad for $b_1$, $b_2$ and
$b_3$, with $b_4$ decoupled from these. We can clearly see that there is no resonant behaviour
in mode $b_4$. 
 \\

We now transform the system to the next order by eliminating non-resonant quartets. The transformation is
\begin{equation}
  \begin{array}{l}
  B_1 = C_1 -\dfrac{z_1 \left(C_4 C_{-2}^2 s_3+C_{-4} C_3^2 s_2\right)}{\left(\omega _2+ \omega
   _3-\omega _4\right){}^2} \\
  B_2 = C_2 +\dfrac{C_{-2} C_{-1} C_4 \left(s_2 z_3-s_3 z_2\right)}{\left(\omega _2+ \omega
   _3-\omega _4\right){}^2} \\
  B_3 = C_3  +\dfrac{C_{-3} C_1 C_4 \left(s_3 z_2-s_2 z_3\right)}{\left(\omega _2+ \omega _3-\omega _4\right){}^2} \\
  B_4 = C_4 +\dfrac{s_4 \left(C_1 C_2^2 z_3+C_{-1} C_3^2 z_2\right)}{\left(\omega _2+ \omega
   _3-\omega _4\right){}^2} \\
  \end{array}
\end{equation}

and the new equations of motion are:
\begin{equation}
  \begin{array}{l}
  \dot{C}_{1}=-i \omega_{1}C_{1} + z_1 C_{2}^*C_{3}+ \mathcal{O}(|\mathbf{C}|^4) \quad \\
  \dot{C}_{2}=-i \omega_{2}C_{2} + z_2 C_{1}^*C_{3} + \dfrac{i s_2 C_2}{\omega_2+\omega_3-\omega_4}\left(-s_4|C_3|^2+s_3|C_4|^2 \right) + \mathcal{O}(|\mathbf{C}|^4)\\
  \dot{C}_{3}=-i \omega_{3}C_{3} + z_3 C_{1}C_{2} + \dfrac{i s_3 C_3}{\omega_2+\omega_3-\omega_4}\left(-s_4|C_2|^2+s_2|C_4|^2 \right) + \mathcal{O}(|\mathbf{C}|^4)\\
  \dot{C}_{4}=-i \omega_{4}C_{4} + \dfrac{i s_4 C_4}{\omega_2+\omega_3-\omega_4}\left(s_3|C_2|^2+s_2|C_3|^2 \right) + \mathcal{O}(|\mathbf{C}|^4)\\
\end{array}
\end{equation}

If we change to interaction representation variables via $ c_j = C_j e^{i \omega_{j} t}, \,\, j=1, \ldots, 4$, we obtain the system
\begin{align*}
  \dot{c}_{1}&=  z_1 c_{2}^*c_{3} + \mathcal{O}(|\mathbf{c}|^4)\\
  \dot{c}_{2}&=  z_2 c_{1}^*c_{3} + \frac{i s_2 c_2}{\omega_2+\omega_3-\omega_4}\left(-s_4|c_3|^2+s_3|c_4|^2 \right) + \mathcal{O}(|\mathbf{c}|^4)\\
  \dot{c}_{3}&=  z_3 c_{1}c_{2} + \frac{i s_3 c_3}{\omega_2+\omega_3-\omega_4}\left(-s_4|c_2|^2+s_2|c_4|^2 \right) + \mathcal{O}(|\mathbf{c}|^4)\\
  \dot{c}_{4}&= \frac{i s_4 c_4}{\omega_2+\omega_3-\omega_4}\left(s_3|c_2|^2+s_2|c_3|^2 \right) + \mathcal{O}(|\mathbf{c}|^4)\\
\end{align*}

Discarding the error terms in the above equations of motion, we easily see that $|c_4|$ is constant for all $t$. Also, the evolution of modes $c_1$, $c_2$, and $c_3$ does not depend on the phase of
$c_4$. Therefore $c_4$ does not contribute to the dynamics of the system, and can
be found by quadrature \textit{a posteriori}. Because of this we can reduce the dimension of
the above dynamical system to 4 variables: $c_1$, $c_2$, $c_3$ and $\arg(c_1 c_2 c_3^*)$.
If we can find three independent first integrals of motion, we can then integrate the system.\\

As we know, the isolated triad is integrable. We have reduced our system to the
isolated triad with quadratic nonlinear corrections to the frequencies corresponding to the modes $c_2$ and $c_3$. These corrections do not change the energies of the individual modes $c_2$ and $c_3$.  Therefore, we would expect to find constants of motion that depend quadratically on the amplitudes, similar to the known ``Manley-Rowe'' invariants for the isolated triad.  In fact, by direct inspection we obtain two quadratic invariants:


\begin{equation*}
  \begin{array}{l}
    I_2 = z_1|c_2|^2-z_2|c_1|^2\,,\\
    I_3 = z_1|c_3|^2-z_3|c_1|^2\,.\\
  \end{array}
\end{equation*}

The third and final constant of motion is more difficult to find. After basing our guess on the Hamiltonian of the isolated triad and some trial and error, the final integral was found to be:

$$I_4 = \Im (c_1 c_2 c_3^*) + \frac{s_4}{4\omega^4_{23}}\left(\frac{s_3 }{z_2}|c_2|^4-\frac{s_2 }{z_3}|c_3|^4 \right).$$

Thus our system is integrable. We can reduce our system into a one dimensional potential equation. Define the variable 
$$x(t) = |c_1|^2\,.$$
Then,

\begin{eqnarray}
  \frac{\mathrm{d}x}{\mathrm{d}t} &=& 2 \Re(c_1^* \dot{c}_1) = 2 z_1 \Re(c_1c_2 c_3^*)\nonumber \\ 
  \Longrightarrow  \left(\frac{\mathrm{d}x}{\mathrm{d}t}\right)^2 &=& 4 z_1^2 [\Re(c_1c_2 c_3^*)]^2 = 4 z_1^2 \left(|c_1|^2 |c_2|^2 |c_3|^2 - [\Im(c_1c_2 c_3^*)]^2\right),\nonumber
\end{eqnarray}

\begin{eqnarray}
  \left(\frac{\mathrm{d}x}{\mathrm{d}t}\right)^2 &=& 4 x(I_2+z_2 x)(I_3 +z_3 x)\nonumber\\ & -& 4z_1^2\left(I_4 + \frac{s_4}{4 \omega^4_{23}}\left(\frac{s_2}{z_1^2z_3}(I_3+z_3 x)^4 - \frac{s_3}{z_1^2z_2}(I_2+z_2 x)^4 \right) \right)^2\,.\nonumber
\end{eqnarray}

  \begin{figure}
     \centering
       \includegraphics[width=0.5\textwidth]{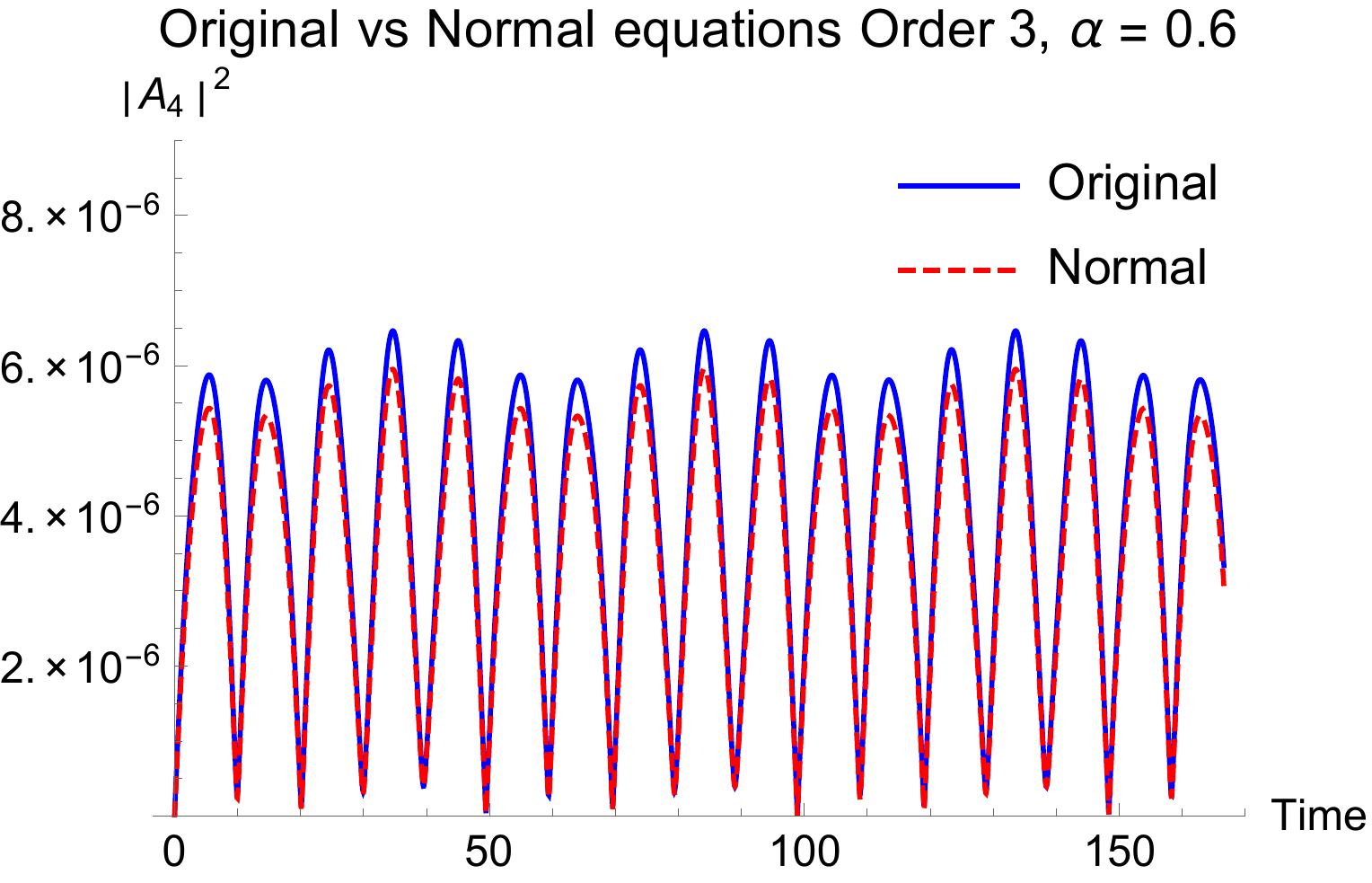}
       \includegraphics[width=0.45\textwidth]{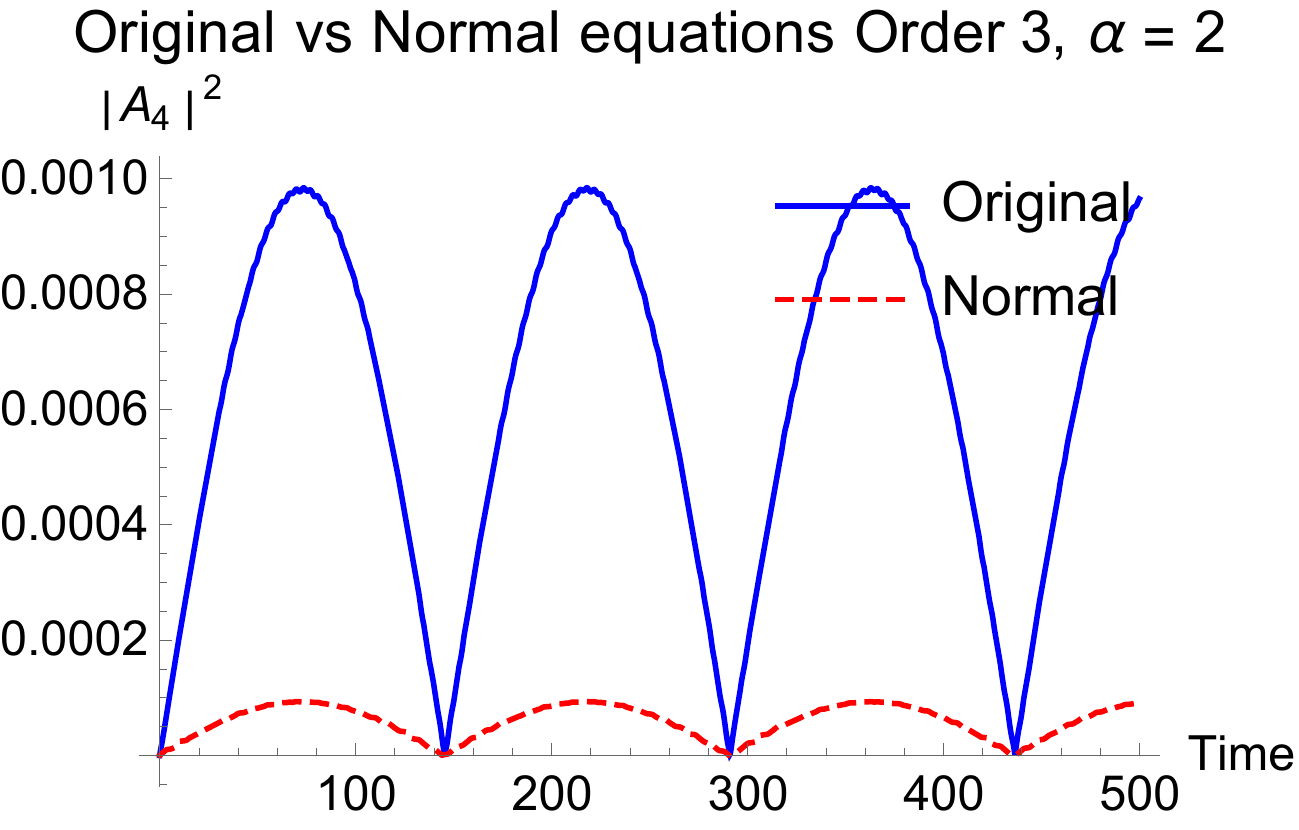}
   \caption{Left panel: Comparison of $|A_4|^2$ calculated from the original equations and the transformed equations with $\alpha<\alpha_r$.
     Right panel: Comparison of $|A_4|^2$ calculated from the original equations and the transformed equations with $\alpha\approx\alpha_r$.
 }
   \label{fig:nearconv}
   \end{figure}

Comparing the solution of the resonant quartet normal equations transformed back to our 
original variables shows wildly different behaviour to the numerical solution to the original
equations near the point of resonance as can be seen in figure~\ref{fig:nearconv}. As our transformation is a power series, this suggests
that there is an issue of convergence with the transformation around the point of resonance.

\section{Convergence of the transformation}
\label{sec:pwc}
In classical wave turbulence theory a weakly nonlinear
regime with infinitesimal amplitudes is considered so the system should be well inside the 
region of convergence. However in order to trigger the zero-th harmonic of precession resonance we 
require the linear timescale to be comparable with the nonlinear timescale, so consideration
of the convergence and applicability of the transformation is required. To study the 
convergence we look at the relative sizes of the terms in the expansion. We can express the 
normal form transformation as:

$$\mathbf{B} = \mathbf{A} + \mathbf{G}^{(2)}(\mathbf{A}) + \mathbf{G}^{(3)}(\mathbf{A}) + \ldots\,,$$
where $\mathbf{A} = (A_1,A_2,A_3,A_4)^T$, $\mathbf{B} = (B_1,B_2,B_3,B_4)^T$ and 
$\mathbf{G}^{(n)}(\mathbf{A})$
is a vector whose components are monomials of degree $n$ of the components of $\mathbf{A}$.


To quantify the rate of convergence of the power series we first need to consider how small
our terms become with increasing order within the domain of convergence. 
As the normal form transformation is a power series, we expect that in the domain of 
convergence the size of $\mathbf{G}^{(n)}$ decreases exponentially with respect to increasing order, i.e.

$$\|{\mathbf{G}^{(n)}}\| \sim e^{\lambda n}\,,$$
where $\lambda$ is our rate of convergence and $\|\cdot\|$ denotes an appropriately defined norm.  As our $\mathbf{G}^{(n)}$'s are vector valued functions, there is a certain amount of
arbitrariness to how we choose our norm to determine the size of our terms. As our 
reduced system is based on a physical system, it makes sense to use a norm based on a
physical quantity. In our case we choose to base our norm in terms of the energy of the system,
$E = \sum_{i=1}^4(|\mathbf{k}_i|^2+F)|A_i|^2 $, i.e. $\|\mathbf{x}\|^2=\sum_{i=1}^4(|\mathbf{k}_i|^2+F)|x_i|^2.$\\

Within the domain of convergence $\lambda$ should
be negative.
We wish to study the convergence of the normal form transformation at a set of initial 
conditions and along a trajectory for specific initial conditions. We use the ratio test to 
determine whether the series converges or not. We do this numerically by performing a linear
regression on $\log\left(\|{\mathbf{G}^{(n)}}\|\right)$ as a function of $n$. We choose to look
at $\log\left(\|{\mathbf{G}^{(n)}}\|\right)$ because our terms decrease in size exponentially as a function of
$n$ in the region of convergence
. The slope of our regression line corresponds to $\lambda$, our rate of
convergence.  \\

   \begin{figure}
     \centering
       \includegraphics[width=0.5\textwidth]{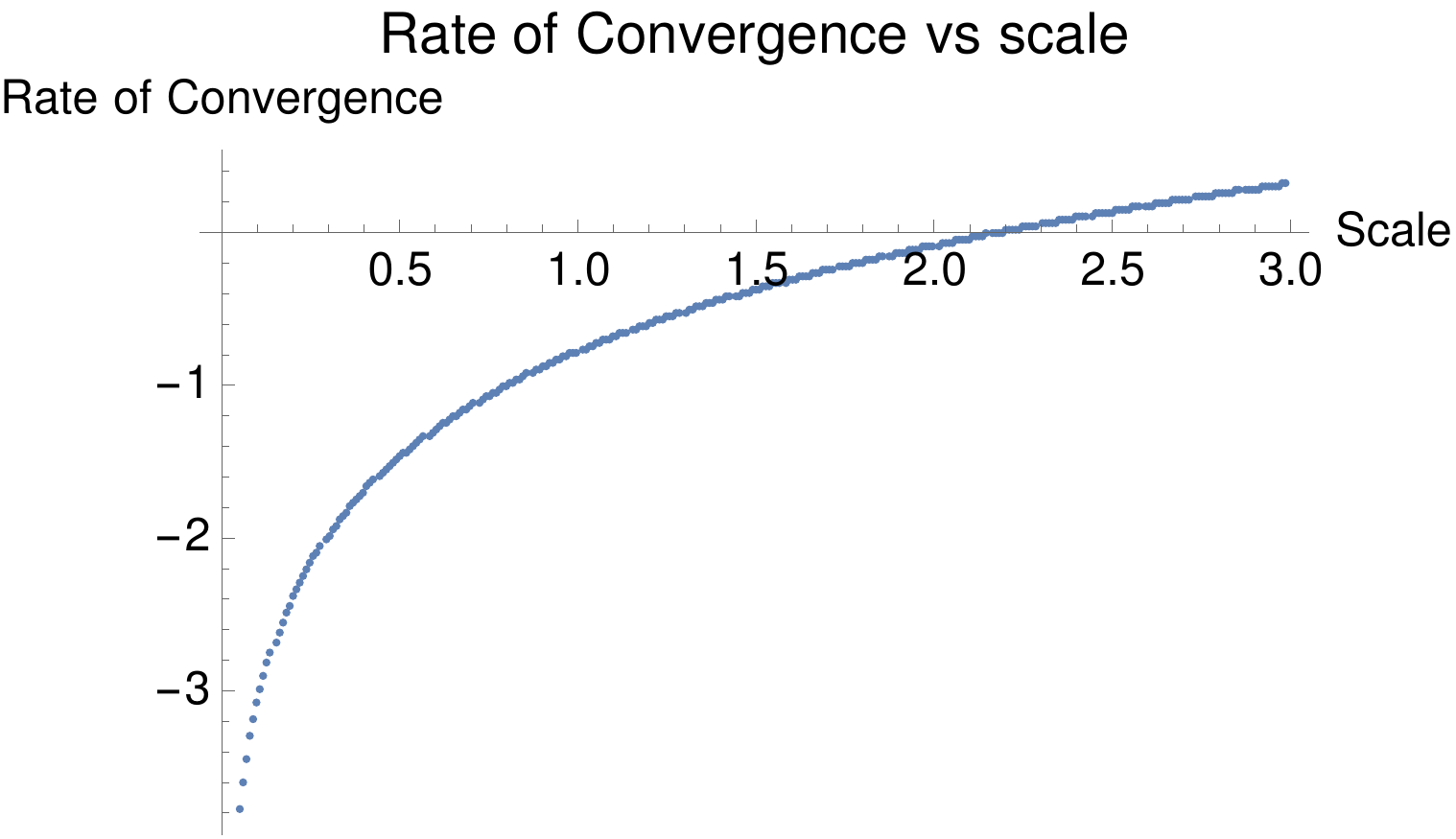}
       \includegraphics[width=0.45\textwidth]{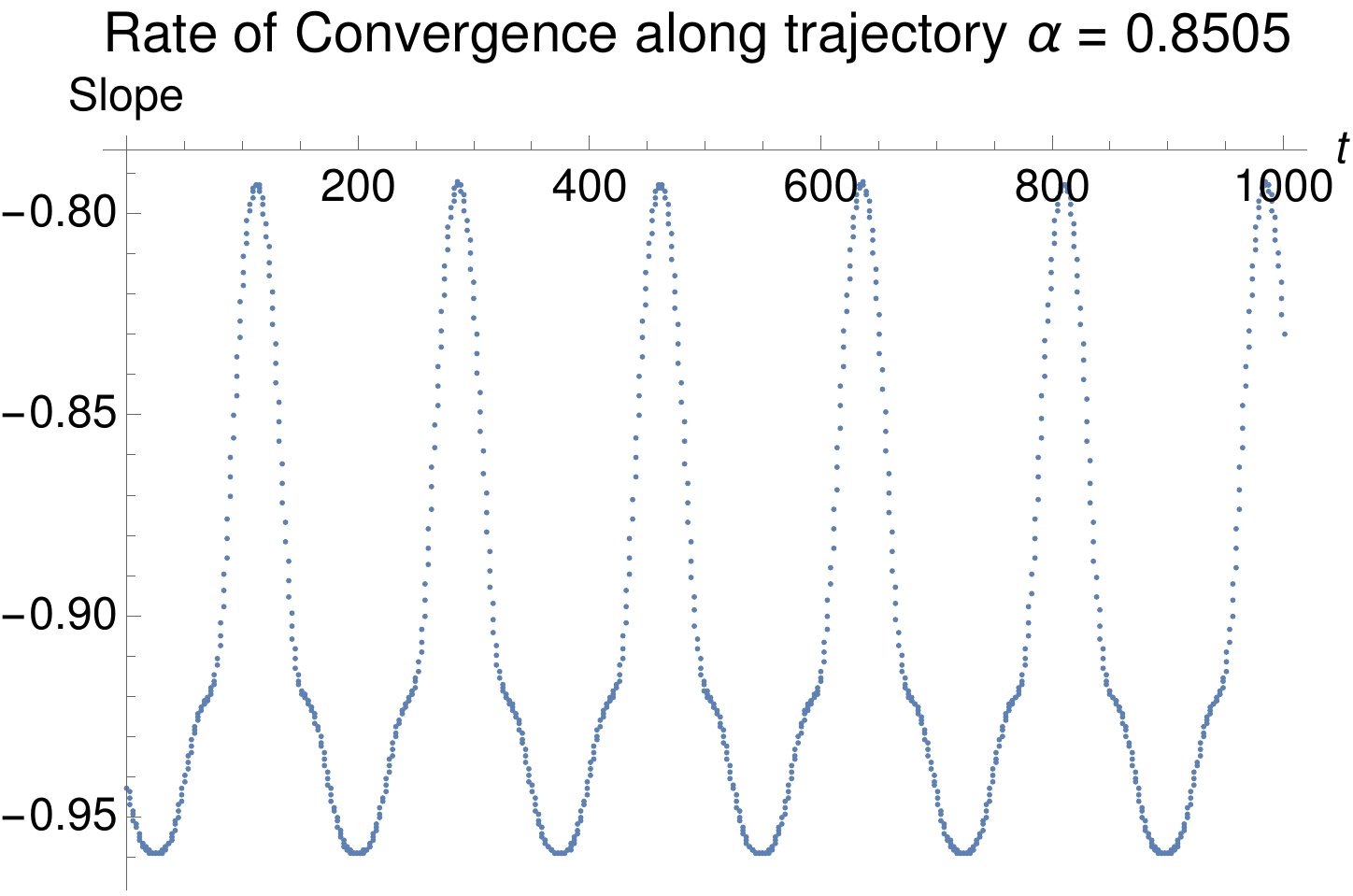}
   \caption{Left panel: Rate of exponential convergence of the normal-form transformation as a function of the initial
   condition's scaling parameter $\alpha$. 
   Right panel: Exponential rate of
 convergence for a given trajectory in time. Here we use the same initial conditions as 
 figure~\ref{fig:4mode}.}
   \label{fig:pwconv1}
   \end{figure}

In figure~\ref{fig:pwconv1} we can see that by increasing the scaling factor $\alpha$ in front of
our amplitudes, taken from equation (\ref{eq:init_conds}), our rate of exponential convergence becomes slower and slower until
at  $\alpha \approx 2.18$ the transformation begins to diverge. This is very close to the 
scaling value required for precession resonance we found in Section~\ref{sec:intro}. 
On the right we can see that, along the trajectory obtained from the initial condition given by equation (\ref{eq:init_conds}) with $\alpha=0.8505$, the calculated rate of convergence
oscillates, although it does not change drastically over the course of the whole run. This will
undoubtedly introduce some uncertainty in calculating our value for the rate of convergence 
for a given trajectory, however for what we want to say it is only necessary to have
an approximate idea of the region of convergence, so using the calculated rate of convergence
at our chosen initial condition is sufficient.

   \begin{figure}
     \centering
       \includegraphics[width=0.44\textwidth]{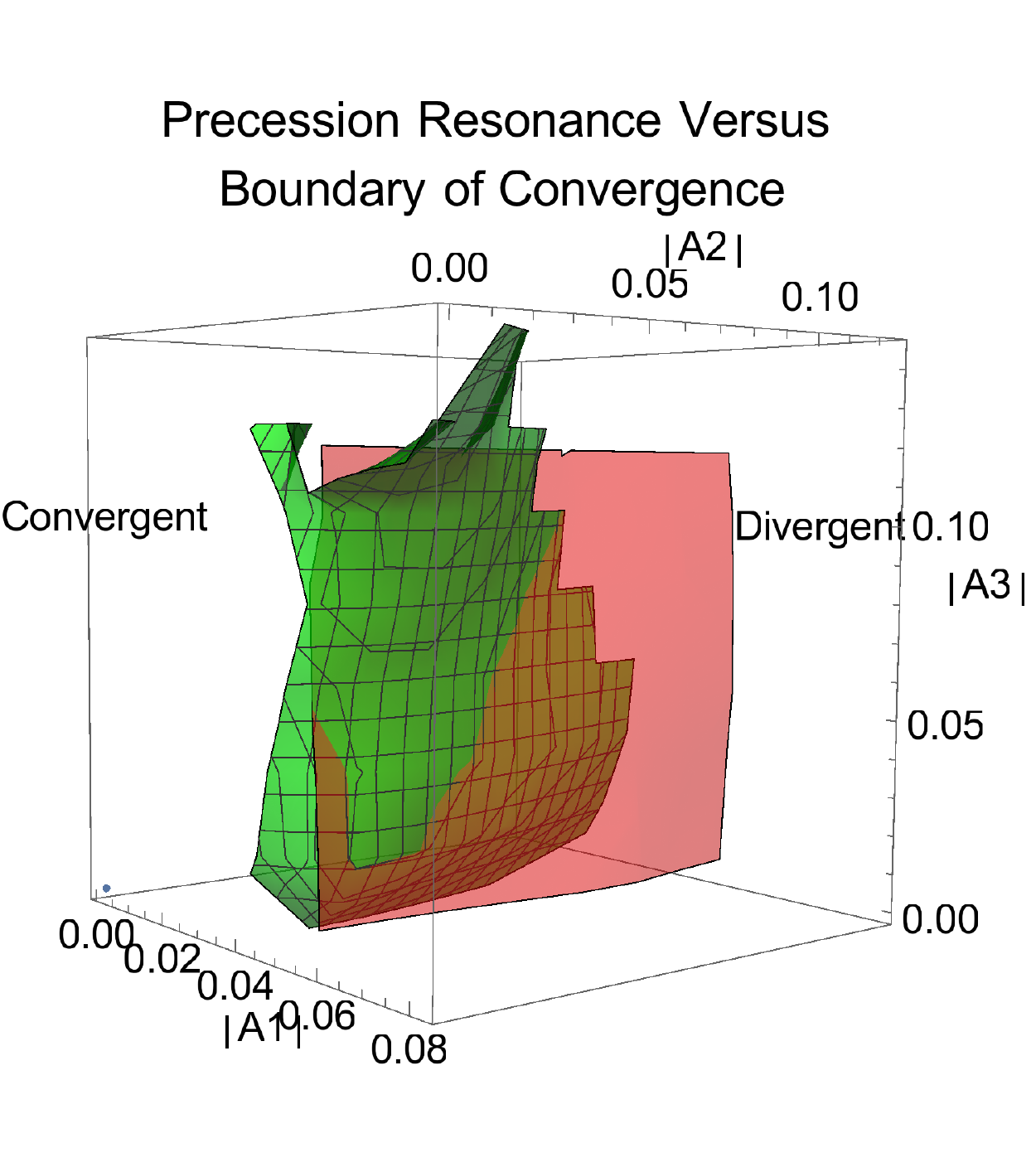}
       \includegraphics[width=0.44\textwidth]{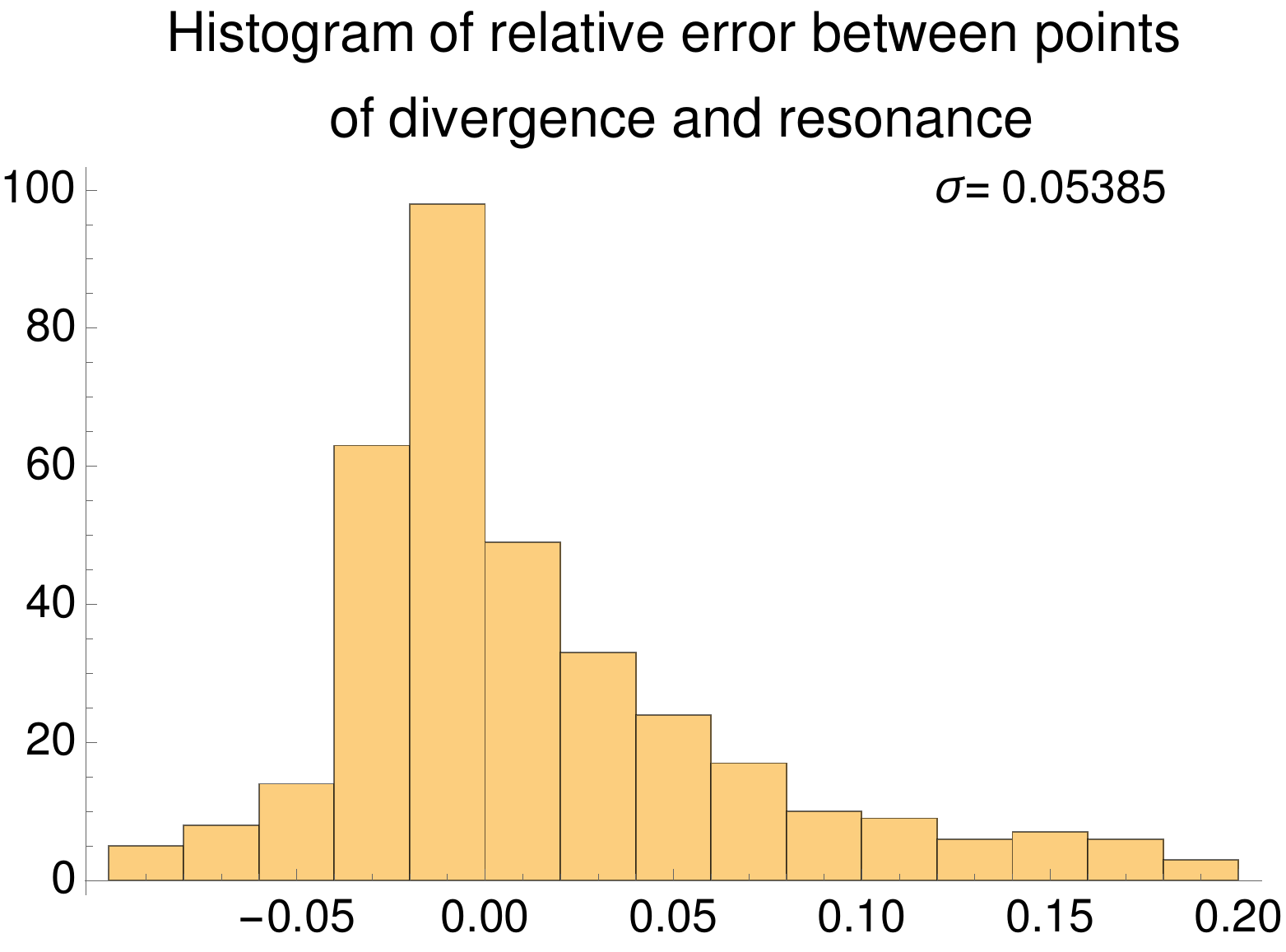}
     \caption{(Colour online). Left panel: Boundary of the region of convergence for the 
     normal form transformation (green meshed opaque surface), and set of initial conditions which lead to
     precession resonance (red unmeshed semi-transparent surface). Right panel: Distribution of the relative difference between
 the manifold of precession resonance and the boundary of convergence.  }
   \label{fig:setconv}
   \end{figure}

Extending this to more general initial conditions, or more specifically to more general points in state space, we wish to compare the region of state space where  
the normal form transformation diverges, with the set of points where precession resonance occurs, by varying $|A_1|$, $|A_2|$ 
and $|A_3|$ while keeping $|A_4(0)|=0$ and keeping the same phases as calculated from equation (\ref{eq:init_conds}). Since we are 
investigating precession resonance to $A_4$, we perform a search over the state-space region where $|A_1|>|A_2|$ and  $|A_1|>|A_3|$ so
that the energy transfer from $A_1$ to $A_4$ is favoured.
The search for precession resonance is done along rays in the $|A_1|, |A_2|, |A_3|$ space, i.e.~we take a representative  point on the ray as initial condition and then rescale uniformly this initial condition until a precession resonance is found, which allows us to calculate the initial condition along the ray which leads to precession resonance. We then compare this to the point along the ray
at which the rate of convergence becomes zero. The result is shown in
 figure~\ref{fig:setconv}: in the left panel, the green meshed surface marks the boundary between
the region of convergence and divergence and the red semi-transparent surface shows the set of initial conditions that lead to precession resonance in the original dynamical system. At a first
glance it seems that these two surfaces are qualitatively different; however, the important
fact to note is that they are quite close, and at many points precession resonance
occurs within the region of divergence.
On the right panel of the figure we show the histogram of the relative difference between the scaling for
divergence of the normal form transformation and the scaling for precession resonance. There is a strong peak around 0 with a standard deviation of $\sigma = 0.05$, suggesting a strong correlation
between precession resonance and the divergence of the transformation.

   \begin{figure}
     \centering
       \includegraphics[width=0.5\textwidth]{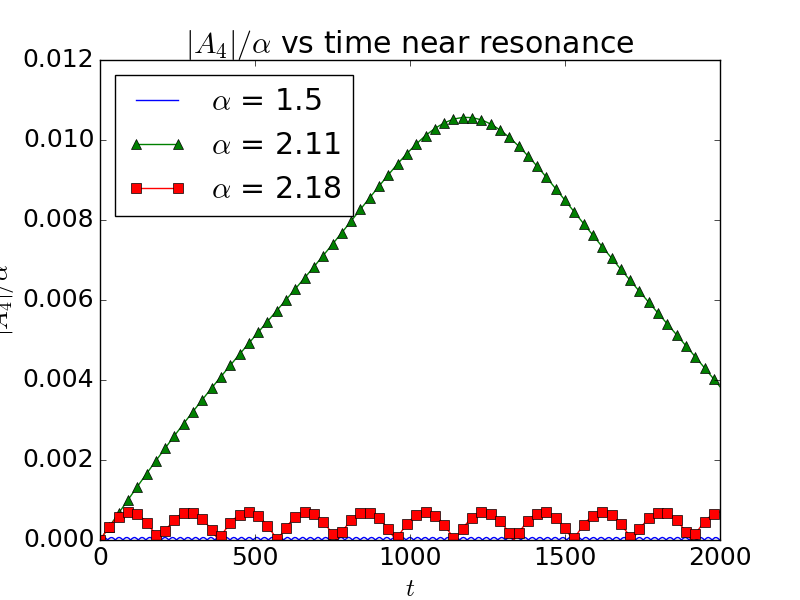}
       \includegraphics[width=0.45\textwidth]{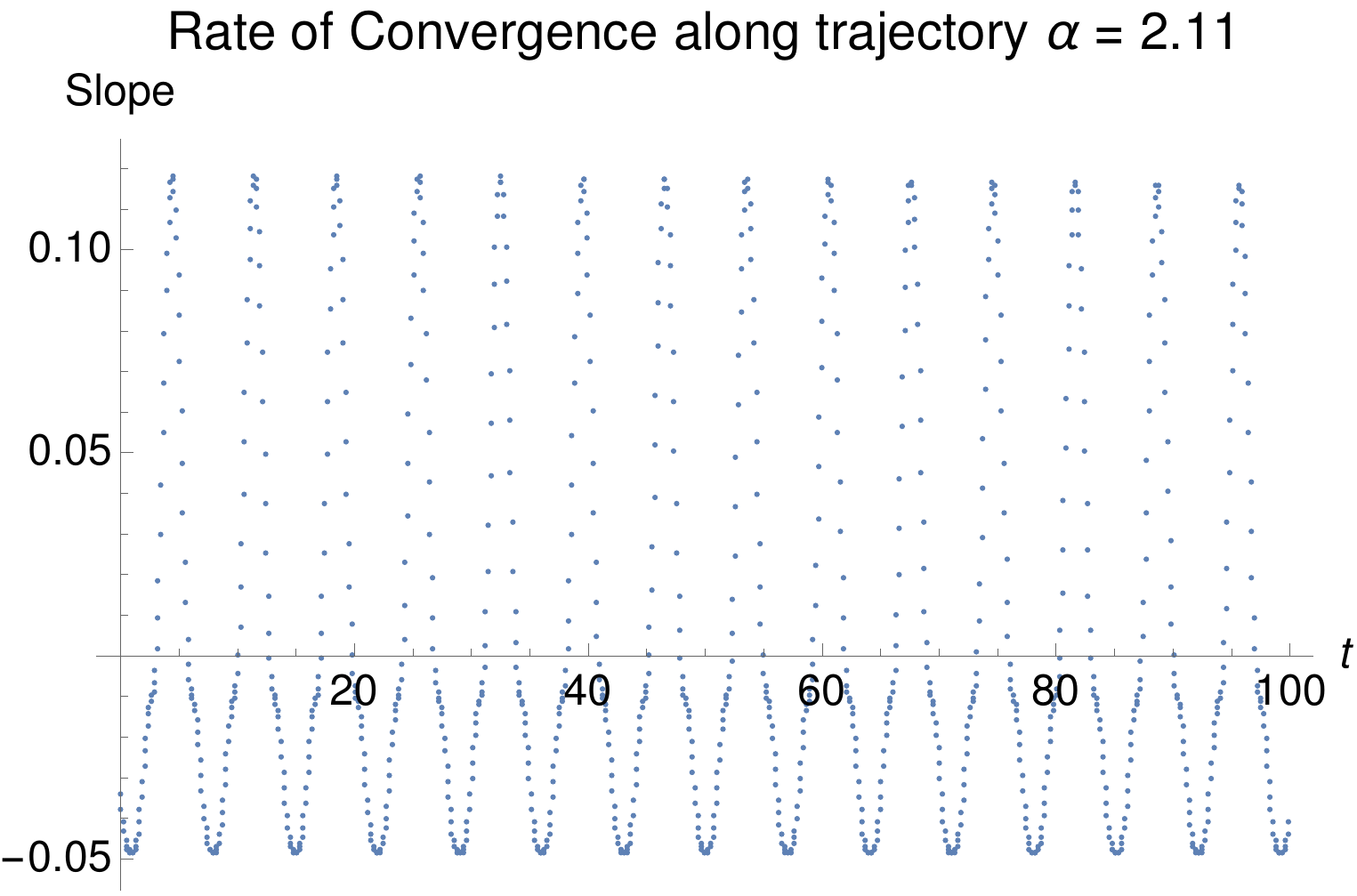}
     \caption{Left panel: For a given initial condition of the form of equation (\ref{eq:init_conds}), the point of resonance ($\alpha \approx 2.11$) is close to the 
       calculated point of divergence ($\alpha \approx 2.18$), but
       is still within the calculated region of convergence.
       Right panel: Even though the initial condition from equation (\ref{eq:init_conds}) is within
   our calculated region of convergence, the actual rate of convergence at the initial condition is very slow, and moreover the actual trajectory repeatedly enters the region of divergence.}
   \label{fig:icconv}
   \end{figure}
Seeing that precession
resonance occurs quite often at or outside the boundary of convergence of the normal
form transformation, we can infer that the resonance cannot be captured by the 
transformation at these points. However consideration is needed for the points on the 
manifold of resonance which are inside the region of convergence. Firstly as we have seen 
from figure~\ref{fig:pwconv1} our calculated rate of convergence varies somewhat over the 
trajectory. If we take our initial conditions from Section~\ref{sec:intro} we can
seen that precession resonance occurs within our calculated region of convergence from
figure~\ref{fig:icconv} for these particular initial conditions. 
However the rate of convergence at this initial value is very slow, and at points of the trajectory the transformation becomes divergent. 
So if this initial condition which leads to precession resonance is in the region of convergence of the transformation, it would require many terms to be accurately described. 
The fastest convergence we have on this trajectory is $\lambda=-0.05$, which would require
computing the transformation to order  $n\approx 46$ for our error to be $10\%$.  
To check the feasibility of calculating the normal form transformation up to order $n=46$, we 
count the number of terms that on the right hand side of the evolution equations at each order
and then perform linear regression on the $\log$ of the number of terms. We found that 
the number of terms at order $n$ $\sim e^{1.01 n}$ , which would imply that $n=46$ would have of the order of $10^{20}$ terms, which would be unfeasable to calculate.
Because it is computationally difficult to compute the transformation for orders higher than $4$ for a full system, it would be impossible for precession resonance to manifest itself in normal coordinates in the usual theory.

\section{Conclusions}
\label{sec:concl}
In this paper we study the relationship between on the one hand the normal form transformations used in 
the classical theory of wave turbulence, and on the other hand precession resonance, a finite-amplitude 
phenomenon characterised by strong energy transfers first introduced in~\citep{Busta}.
It is already known that the normal form transformation is well behaved in regimes where the nonlinear 
frequencies are far smaller than the linear frequencies,
but this study allows us to understand how and when the normal form transformation breaks down as the
amplitudes are set beyond the weakly nonlinear limit.
We made this study more tractable by classifying the manifolds associated with precession 
resonance in a 4-mode reduced model of the CHM equation. In our reduced model, 
there appears to be a strong relationship between the amplitudes at which precession resonance
occurs and the points at which the normal form transformation diverges.
Although the rate of convergence varying over a trajectory means there will be a certain
amount of uncertainty in choosing the exact scaling which causes divergence in the 
transformation, in fact it is not fully necessary to pinpoint exactly the region of convergence of the transformation in order to conclude that the scales at which precession resonance occurs are too large 
to be captured in the transformation. This notion makes intuitive sense, as precession resonance occurs when the linear timescales
become commensurate with the characteristic nonlinear timescales of the system, which would
suggest that the nonlinear terms in the evolution equations would become commensurate with
the linear terms, thus suggesting that a power series in these variables would diverge. From a
dynamical systems point of view, our work shows that the region of convergence of the normal form transformation is strongly 
related to the stable manifold associated with precession 
resonance. For more general systems in higher dimensions, knowledge
of the transformation could allow us to know approximate initial conditions which lead to
precession resonance \textit{a priori} and without needing to use the amplitude scaling 
search done in this paper and in~\citep{Busta}. The results of this paper shed some light on 
the dynamics at intermediate nonlinearity, and provide a step further to bridge the gap between strong and
weak nonlinearity.

~\\[.5cm]

\bibliographystyle{jfm}

\bibliography{normal_form_ref}

\appendix
\section{Higher order terms in the expansion}

Recall that we defined $\mathbf{B}=(B_1,B_2,B_3,B_4)^T$.

In Section \ref{sec:transf} we defined the transformed evolution equations in vector notation
as:

$$\mathbf{\dot{B}} = J \mathbf{B} + R^{(2)}(\mathbf{B}) + R^{(3)}(\mathbf{B}) + \ldots$$

Component-wise, the equations look like:

$$\dot{B_j} = -i \omega_j B_j + R^{(2)}_j(\mathbf{B}) + R^{(3)}_j(\mathbf{B}) + \ldots$$

\begin{align*}
  J=\left(\begin{array}{cccc}
    -i\omega_1 & 0 & 0 & 0 \\ 
    0 & -i\omega_2 & 0 & 0 \\ 
    0 & 0 & -i\omega_3 & 0 \\ 
    0 & 0 & 0 & -i\omega_4 \\ 
\end{array}\right)
\end{align*}

\begin{align*}
  R^{(2)}_1(\mathbf{B})&= B_{2}^* B_3 z_1 \\
  R^{(2)}_2(\mathbf{B})&=B_{1}^* B_3 z_2 \\
  R^{(2)}_3(\mathbf{B})&=B_1 B_2 z_3 \\
  R^{(2)}_4(\mathbf{B})&=0 \\
 \end{align*}

\begin{align*}
  R^{(3)}_1(\mathbf{B})&=0 \\
 R^{(3)}_2(\mathbf{B})&=
 \frac{i B_2 s_2 \left(B_{-4} B_4 s_3+B_{-3} B_3 s_4\right)}{\omega _1+2 \omega
   _2-\omega _4} \\
 R^{(3)}_3(\mathbf{B})&=
 \frac{i B_3 s_3 \left(B_{-4} B_4 s_2+B_{-2} B_2 s_4\right)}{\omega _1+2 \omega
   _2-\omega _4} \\
 R^{(3)}_4(\mathbf{B})&=
 -\frac{i B_4 s_4 \left(B_{-3} B_3 s_2+B_{-2} B_2 s_3\right)}{\omega _1+2 \omega
   _2-\omega _4} \\
 \end{align*}

\begin{align*}
  F^{(4)}_1(\mathbf{B})&=\frac{B_{4}^* B_{2}^* B_3 B_4 s_2 s_3 z_1}{\left(\omega _1+2 \omega _2-\omega
   _4\right){}^2} \\
   F^{(4)}_2(\mathbf{B})&=\frac{s_2 \left(B_{3}^* s_4 \left(B_1 B_2^2 z_3+B_{1}^* B_3^2 z_2\right)+B_{4}^* B_{1}^*
   B_3 B_4 \left(s_3 z_2-s_2 z_3\right)\right)}{\left(\omega _1+2 \omega _2-\omega
   _4\right){}^2} \\
   F^{(4)}_3(\mathbf{B})&=\frac{s_3 \left(B_{2}^* s_4 \left(B_1 B_2^2 z_3+B_{1}^* B_3^2 z_2\right)+B_{4}^* B_1 B_2
   B_4 \left(s_2 z_3-s_3 z_2\right)\right)}{\left(\omega _1+2 \omega _2-\omega
   _4\right){}^2} \\
   F^{(4)}_4(\mathbf{B})&=\frac{\left(B_{3}^* B_1 B_2-B_{2}^* B_{1}^* B_3\right) B_4 s_4 \left(s_3 z_2-s_2
   z_3\right)}{\left(\omega _1+2 \omega _2-\omega _4\right){}^2} \\
 \end{align*}

 The method used for eliminating non-resonant terms is done order by order, so to get the full
 transformation from the original variables we need to compose the transformation for each 
 order. Below $\mathbf{A}$ is a vector of our original variables and $\mathbf{B}$ are our
 normal variables at the order we desire.

 \begin{equation*}
   \begin{split}
     \mathbf{A} &= \mathbf{B} + h^{(2)}(\mathbf{B}) + h^{(3)}(\mathbf{B}) + \ldots \\
     A_j &= B_j + h^{(2)}_j(\mathbf{B}) + h^{(3)}_j(\mathbf{B}) + \ldots \\
   \end{split}
 \end{equation*}

\begin{align*}
 h^{(2)}_1(\mathbf{B})&= 0 \\
 h^{(2)}_2(\mathbf{B})&=-\frac{i B_{3}^* B_4 s_2}{\omega _1+ \omega _2-\omega _3} \\
 h^{(2)}_3(\mathbf{B})&=-\frac{i B_{2}^* B_4 s_3}{\omega _1+ \omega _2-\omega _3} \\
 h^{(2)}_4(\mathbf{B})&=\frac{i B_2 B_3 s_4}{\omega _1+ \omega _2-\omega _3}  
 \end{align*}

\begin{align*}
  h^{(3)}_1(\mathbf{B})&= 
  -\frac{z_1 \left(B_4 B_{2}^{*2} s_3+B_{4}^* B_3^2 s_2\right)}{\left(\omega _1+2 \omega
   _2-\omega _4\right){}^2} \\
 h^{(3)}_2(\mathbf{B})&=
 \frac{B_{2}^* B_{1}^* B_4 \left(s_2 z_3-s_3 z_2\right)}{\left(\omega _1+2 \omega
   _2-\omega _4\right){}^2} \\
   h^{(3)}_3(\mathbf{B})&=
 \frac{B_{3}^* B_1 B_4 \left(s_3 z_2-s_2 z_3\right)}{\left(\omega _1+2 \omega _2-\omega
   _4\right){}^2} \\
 h^{(3)}_4(\mathbf{B})&=
 \frac{s_4 \left(B_1 B_2^2 z_3+B_{1}^* B_3^2 z_2\right)}{\left(\omega _1+2 \omega
   _2-\omega _4\right){}^2} \\
 \end{align*}

 \begin{align*}
h^{(4)}_1(\mathbf{B})&= \frac{i B_1 z_1 \left(B_{3}^* B_{2}^* B_4 \left(s_2 z_3-3 s_3 z_2\right)+B_{4}^* B_2 B_3
   \left(3 s_2 z_3-s_3 z_2\right)\right)}{\left(\omega _1+2 \omega _2-\omega
   _4\right){}^3} \\
 h^{(4)}_2(\mathbf{B})&= -\frac{i \left(B_{4}^* B_3 B_2^2 s_3 \left(s_2 s_4+z_1 z_2\right)+B_{1}^* B_3 s_2
     \left(B_{3}^* B_3 s_4 z_2+B_{4}^* B_4 \left(s_3 z_2-s_2 z_3\right)\right)\right)}{\left(\omega _1+2 \omega _2-\omega _4\right){}^3}\\&+\frac{\left(B_{3}^* B_1
     \left(B_2^2 s_2 s_4 z_3+2 B_{1}^* B_4 z_2 \left(s_3 z_2-s_2 z_3\right)\right)\right)}{\left(\omega _1+2 \omega _2-\omega _4\right){}^3}\\&+\frac{\left(B_{3}^*
   B_4 \left(B_{3}^* B_3 s_2 \left(s_2 s_4-z_1 z_2\right)+B_{2}^* B_2 \left(s_3 z_1
   z_2-s_2 z_1 z_3+2 s_2 s_3 s_4\right)\right)+B_{4}^* B_{3}^* B_4^2 s_2^2
   s_3\right)}{\left(\omega _1+2 \omega _2-\omega _4\right){}^3} \\
 h^{(4)}_3(\mathbf{B})&= -\frac{i \left(B_3^2 \left(B_{2}^* B_{1}^* s_3 s_4 z_2+B_{-4} B_2 s_2 \left(s_3 s_4+z_1
       z_3\right)\right)\right)}{\left(\omega _1+2 \omega _2-\omega _4\right){}^3}\\&+\frac{\left(B_1 \left(B_{2}^* B_2^2 s_3 s_4 z_3-B_4 \left(s_3 z_2-s_2 z_3\right)
       \left(B_{4}^* B_2 s_3+2 B_{2}^* B_{1}^* z_3\right)\right)\right)}{\left(\omega _1+2 \omega _2-\omega _4\right){}^3}\\&+\frac{\left(B_{2}^* B_4 \left(B_{2}^* B_2
   s_3 \left(s_3 s_4-z_1 z_3\right)+B_{3}^* B_3 \left(-s_3 z_1 z_2+s_2 z_1 z_3+2 s_2 s_3
   s_4\right)\right)+B_{4}^* B_{2}^* B_4^2 s_2 s_3^2\right)}{\left(\omega _1+2 \omega
   _2-\omega _4\right){}^3} \\
 h^{(4)}_4(\mathbf{B})&= \frac{i s_4 \left(B_{1}^* B_3 \left(B_{2}^* B_4 \left(s_2 z_3-s_3 z_2\right)-4 B_1 B_2
   z_2 z_3\right)+B_{2}^* \left(B_3 B_2^2 \left(s_3 s_4-z_1 z_3\right)+B_{3}^* B_4^2 s_2
   s_3\right)\right)}{\left(\omega _1+2 \omega _2-\omega _4\right){}^3}\\&+\frac{\left(B_2 \left(B_{3}^* \left(B_3^2 \left(s_2 s_4-z_1 z_2\right)+B_1 B_4
   \left(s_3 z_2-s_2 z_3\right)\right)+2 B_{4}^* B_3 B_4 s_2
   s_3\right)\right)}{\left(\omega _1+2 \omega _2-\omega _4\right){}^3} \\
 \end{align*}
The inverse of this transformation which was used to calculate the transformation from the
original variables to the normal variables was defined as:

\begin{equation*}
  \begin{split}
    \mathbf{B} &= \mathbf{A} + G^{(2)}(\mathbf{A}) + G^{(3)}(\mathbf{A}) + \ldots \\
    B_j &= A_j + G^{(2)}_j(\mathbf{A}) + G^{(3)}_j(\mathbf{A}) + \ldots \\
  \end{split}
\end{equation*}

\begin{align*}
 G^{(2)}_1(\mathbf{A})&= 0 \\
 G^{(2)}_2(\mathbf{A})&=\frac{i A_{3}^* A_4 s_2}{\omega _1+ \omega _2-\omega _3} \\
 G^{(2)}_3(\mathbf{A})&=\frac{i A_{2}^* A_4 s_3}{\omega _1+ \omega _2-\omega _3} \\
 G^{(2)}_4(\mathbf{A})&=-\frac{i A_2 A_3 s_4}{\omega _1+ \omega _2-\omega _3}  
 \end{align*}

 \begin{align*}
   G^{(3)}_1(\mathbf{A})&= \frac{z_1 \left(A_4 A_{2}^{*2} s_3+A_{4}^* A_3^2 s_2\right)}{\left(\omega _1+2 \omega _2-\omega _4\right){}^2} \\
 G^{(3)}_2(\mathbf{A})&= \frac{A_{2}^* A_{1}^* A_4 \left(s_3 z_2-s_2 z_3\right)+A_2 s_2 \left(A_{4}^* A_4 s_3+A_{3}^* A_3 s_4\right)}{\left(\omega _1+2 \omega _2-\omega _4\right){}^2} \\
 G^{(3)}_3(\mathbf{A})&= \frac{A_4 \left(A_{3}^* A_1 \left(s_2 z_3-s_3 z_2\right)+A_{4}^* A_3 s_2 s_3\right)+A_{2}^* A_2 A_3 s_3 s_4}{\left(\omega _1+2 \omega _2-\omega _4\right){}^2} \\
 G^{(3)}_4(\mathbf{A})&= \frac{s_4 \left(A_4 \left(A_{3}^* A_3 s_2+A_{2}^* A_2 s_3\right)-A_1 A_2^2 z_3-A_{1}^* A_3^2 z_2\right)}{\left(\omega _1+2 \omega _2-\omega _4\right){}^2} \\
\end{align*}

 \begin{align*}
  G^{(4)}_1(\mathbf{A})&= \frac{i z_1 \left(A_{3}^* A_{2}^* \left(A_1 A_4 \left(3 s_3 z_2-s_2 z_3\right)+A_3^2 s_2
      s_4\right)+A_{4}^* A_1 A_2 A_3 \left(s_3 z_2-3 s_2 z_3\right)-A_2 A_3 A_{2}^{*2} s_3
   s_4\right)}{\left(\omega _1+ \omega _2-\omega _3\right){}^3} \\
 G^{(4)}_2(\mathbf{A})&=\frac{i \left(A_{4}^* A_3 A_2^2 s_3 z_1 z_2+A_{3}^* A_{2}^* A_4 A_2 \left(s_3 z_1 z_2-s_2
       z_1 z_3+4 s_2 s_3 s_4\right)\right)}{\left(\omega _1+ \omega _2-\omega _3\right){}^3}\\&+\frac{i\left(A_{3}^* A_4 s_2 \left(A_{3}^* A_3 \left(2 s_2 s_4-z_1
       z_2\right)+2 A_{4}^* A_4 s_2 s_3\right)\right)}{\left(\omega _1+ \omega _2-\omega _3\right){}^3}\\&+\frac{i\left(-A_{1}^* \left(s_3 z_2-s_2 z_3\right) \left(A_4
   \left(A_{4}^* A_3 s_2-2 A_{3}^* A_1 z_2\right)+A_{2}^* A_2 A_3
   s_4\right)\right)}{\left(\omega _1+ \omega _2-\omega _3\right){}^3} \\
   G^{(4)}_3(\mathbf{A})&=\frac{i \left(A_1 \left(s_3 z_2-s_2 z_3\right) \left(A_2 \left(A_{4}^* A_4 s_3+A_{3}^*
         A_3 s_4\right)-2 A_{2}^* A_{1}^* A_4 z_3\right)\right)}{\left(\omega _1+ \omega _2-\omega _3\right){}^3}\\&+\frac{i\left(A_2 \left(A_4 A_{2}^{*2} s_3 \left(2 s_3
       s_4-z_1 z_3\right)+A_{4}^* A_3^2 s_2 z_1 z_3\right)\right)}{\left(\omega _1+ \omega _2-\omega _3\right){}^3}\\&+\frac{i\left(A_{2}^* A_4 \left(A_{3}^* A_3
   \left(-s_3 z_1 z_2+s_2 z_1 z_3+4 s_2 s_3 s_4\right)+2 A_{4}^* A_4 s_2
   s_3^2\right)\right)}{\left(\omega _1+ \omega _2-\omega _3\right){}^3} \\
   G^{(4)}_4(\mathbf{A})&=-\frac{i s_4 \left(2 A_{1}^* A_3 z_2 \left(A_{2}^* A_4 s_3-2 A_1 A_2 z_3\right)\right)}{\left(\omega _1+ \omega _2-\omega _3\right){}^3}\\&+\frac{i\left(A_2
   \left(A_{3}^* \left(A_3^2 \left(2 s_2 s_4-z_1 z_2\right)+2 A_1 A_4 s_2
   z_3\right)+A_{2}^* A_2 A_3 \left(2 s_3 s_4-z_1 z_3\right)+4 A_{4}^* A_3 A_4 s_2
   s_3\right)\right)}{\left(\omega _1+ \omega _2-\omega _3\right){}^3} 
 \end{align*}
\end{document}